\documentclass[lettersize,journal]{IEEEtran}
\usepackage{verbatim}
\usepackage{cite}
\usepackage{comment}
\usepackage{amsmath,amssymb,amsfonts}
\usepackage{amsfonts} 
\usepackage{graphicx}
\usepackage{textcomp}
\usepackage{xcolor}
\usepackage{color}
\usepackage{hyperref}
\usepackage{float}
\usepackage{pifont}
\usepackage{algorithm}
\usepackage{algorithmicx}
\usepackage{algpseudocode}
\usepackage{amsmath}
\usepackage{subfigure}
\usepackage{soul}
\usepackage{colortbl}
\usepackage{multirow}
\usepackage{booktabs}
\usepackage{tabularx}
\usepackage{makecell}
\usepackage{url}
\hypersetup{
  colorlinks,
  citecolor=black,
  linkcolor=black,
  urlcolor=black}
\usepackage{threeparttable}
\usepackage{tikz}
\usetikzlibrary{decorations.pathreplacing,calc}

\def\BibTeX{{\rm B\kern-.05em{\sc i\kern-.025em b}\kern-.08em
    T\kern-.1667em\lower.7ex\hbox{E}\kern-.125emX}}

\def\BibTeX{{\rm B\kern-.05em{\sc i\kern-.025em b}\kern-.08em
    T\kern-.1667em\lower.7ex\hbox{E}\kern-.125emX}}
\begin{document}
\renewcommand\ttdefault{cmtt}
\title{DAS-MP: Enabling High-Quality \underline{M}acro \underline{P}lacement with Enhanced \underline{D}ataflow \underline{A}warenes\underline{s}}

\author{Xiaotian~Zhao,~\IEEEmembership{Graduate Student Member,~IEEE,} Zixuan Li,~\IEEEmembership{Student Member,~IEEE,} Yichen Cai,~\IEEEmembership{Graduate Student Member,~IEEE,} Tianju Wang, Jiayin Chen, Yushan Pan,~\IEEEmembership{Senior Member,~IEEE} and Xinfei Guo,~\IEEEmembership{Senior Member,~IEEE}        
\thanks{This work has been submitted to the IEEE for possible publication. Copyright may be transferred without notice, after which this version may no longer be accessible.}
\thanks{This work was supported in part by the National Science and Technology Major Project (Grant No. 2021ZD0114701), the State Key Laboratory of Integrated Chips and Systems (SKLICS) open fund and a SJTU Explore-X grant.}
\thanks{Xiaotian Zhao, Zixuan Li, Yichen Cai, Tianju Wang, Jiayin Chen and Xinfei Guo are with the University of Michigan – Shanghai Jiao Tong University Joint Institute, Shanghai Jiao Tong University, Shanghai 200240, China (E-mails: \{xiaotian.zhao, cai\_yichen, wangtianju, chenjiayin, xinfei.guo\}@sjtu.edu.cn, lzixuan136@gmail.com).}
\thanks{Yushan Pan is with the Xi'an Jiaotong-Liverpool University, Suzhou 215000, China (E-mail: Yushan.Pan@xjtlu.edu.cn).}
\thanks{Corresponding author: Xinfei Guo.}
}



\maketitle

\begin{abstract}
Dataflow is a critical yet underexplored factor in automatic macro placement, which is becoming increasingly important for developing intelligent design automation techniques that minimize reliance on manual adjustments and reduce design iterations. Existing macro or mixed-size placers with dataflow awareness primarily focus on intrinsic relationships among macros, overlooking the crucial influence of standard cell clusters on macro placement. To address this, we propose DAS-MP, which extracts hidden connections between macros and standard cells and incorporates a series of algorithms to enhance dataflow awareness, integrating them into placement constraints for improved macro placement. To further optimize placement results, we introduce two fine-tuning steps: (1) congestion optimization by taking macro area into consideration, and (2) flipping decisions to determine the optimal macro orientation based on the extracted dataflow information. By integrating enhanced dataflow awareness into placement constraints and applying these fine-tuning steps, the proposed approach achieves an average 7.9\% improvement in half-perimeter wirelength (HPWL) across multiple widely used benchmark designs compared to a state-of-the-art dataflow-aware macro placer. Additionally, it significantly improves congestion, reducing overflow by an average of 82.5\%, and achieves improvements of 36.97\% in Worst Negative Slack (WNS) and 59.44\% in Total Negative Slack (TNS). The approach also maintains efficient runtime throughout the entire placement process, incurring less than a 1.5\% runtime overhead. These results show that the proposed dataflow-driven methodology, combined with the fine-tuning steps, provides an effective foundation for macro placement and can be seamlessly integrated into existing design flows to enhance placement quality.

\end{abstract}

\section{Introduction}
\label{introduction}
\IEEEPARstart{W}{ith} increasing process technology complexity and higher levels of chip integration, System on Chip (SoC) has become the mainstream design model which relies on a large number of IPs (internal or external) to improve the reuse of modules, reducing the cost of repeated development and shortening the design cycle. From a logic function perspective, IPs comprise various hardware modules and circuit components, such as memory cells, analog modules (e.g., ADCs, PLLs), and even entire processor cores. However, from a physical design perspective, these IPs come in the form of macros, which are typically larger, pin-dominant, and have a greater impact on timing, congestion, and design rule checking (DRC) violations than standard cells. The quality of macro placement in the floorplanning phase will directly impact how standard cells are placed and routed in later stages, and ultimately the design's quality-of-results (QoR). However, as design scales continue to grow, the complexity of macro placement has significantly increased, becoming a critical bottleneck that affects both design efficiency and quality. \par
\begin{figure}[t]
    \centering
    \includegraphics[width =1.0 \linewidth]{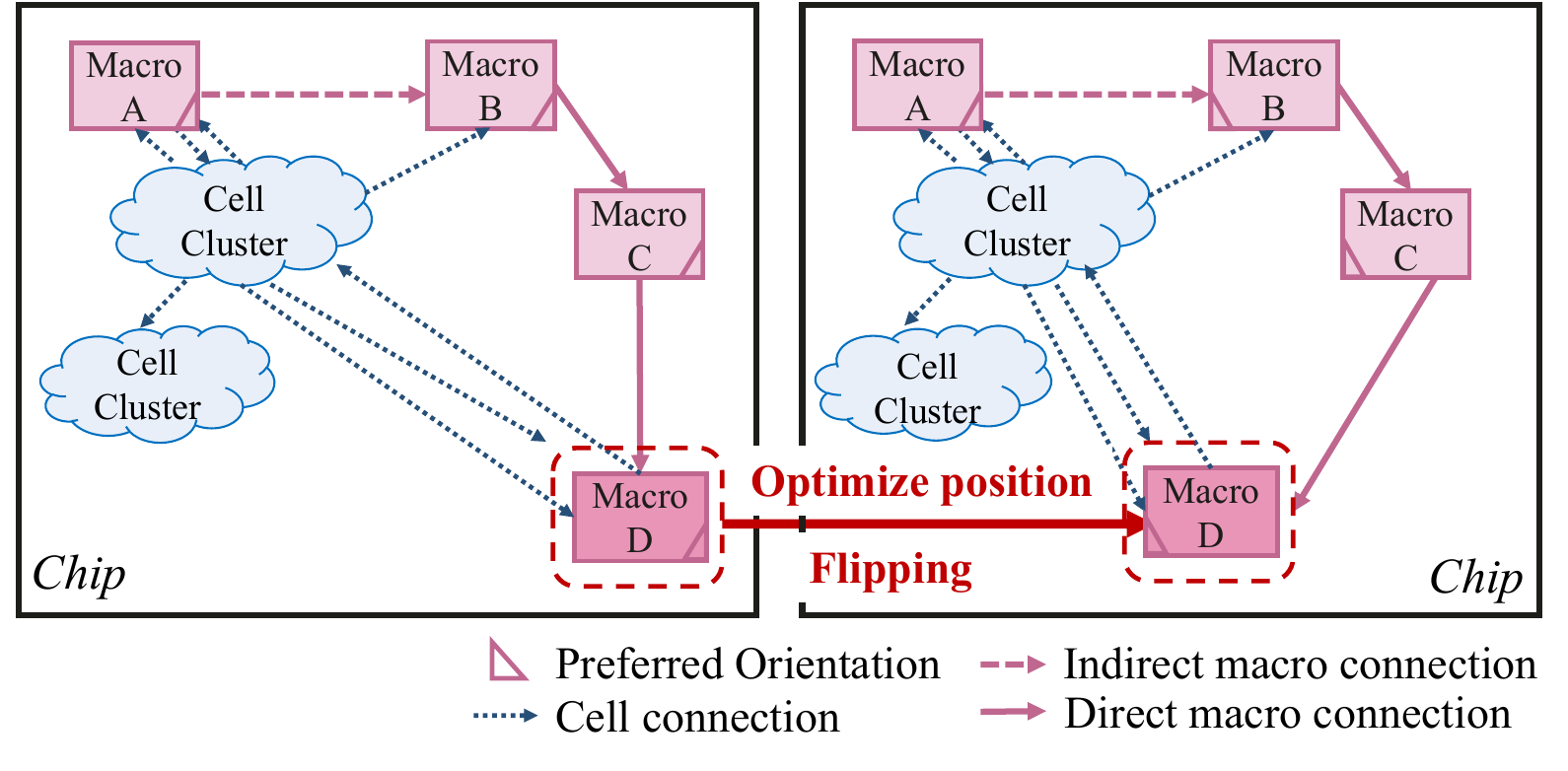}
    \vspace{-15pt}
    \caption{The illustration shows all dataflow connections, with the uncovered dataflow connections indicated by shaded lines.}
    \label{fig:dataflow illustration}
\end{figure}

In the current physical design process, macro placement typically relies on experienced engineers performing manual placements, often requiring multiple iterations to achieve an acceptable placement to some extent. As a result, fully automating macro placement is highly challenging. This is due to the conflict between the very limited design and physical information available in the early stages and the vast search space, making it nearly impossible to accurately predict the timing and other relevant information required for the subsequent placement and routing process (P\&R), thus hindering placement optimization.
Although Reinforcement Learning (RL), Deep Learning (DL), or recently proposed large-language model (LLM)-based methods, such as \cite{Mirhoseini2021, liu2022floorplanning, agnesina2023autodmp,cheng2023assessment}, have shown promise in automating macro placement, they face several significant challenges. First, these methods often struggle with legalization, as the generated placements may violate design rules or physical constraints. Second, the training process for RL/DL/LLM models is computationally intensive, requiring substantial resources and time. Third, obtaining high-quality and diverse training datasets is difficult, especially for complex designs with unique characteristics. Additionally, some of these techniques rely on a good initial placement as a starting point \cite{cheng2023assessment}, which limits their applicability in scenarios where such prior knowledge is unavailable. These limitations highlight the need for complementary techniques to achieve human-quality macro placement results.\par

To automate macro placement and mimic manual strategies, it is crucial to identify the features that physical engineers consider during placement. One prominent feature is dataflow, which engineers use to optimize macro positioning. Dataflow defines the movement of data between macros and standard cells, influencing the circuit’s timing and power consumption. Physical design engineers invest significant effort in analyzing flylines and collaborating with logic design engineers to gain a deep understanding of dataflow information, underscoring the importance of dataflow awareness in macro placement. Since the interactions between macros and other modules can be analyzed from the synthesized netlist. Furthermore, the number of macros is significantly smaller than standard cells in modern designs, making it computationally feasible to extract detailed dataflow connections involving macros. This has inspired dataflow-aware macro placers \cite{Ye2000, Vidal-Obiols2019,lin2021dataflow,Kahng2022,Vidal-Obiols2021,kahng2023hierrtlmp,10546560}, where the dataflow is analyzed and incorporated as constraints to guide macro placement.

Current macro placers or mixed-size placers incorporate some level of dataflow analysis, but they mainly only focus on the virtual or direct connections between macros (i.e. macro-macro connections), as the placement of standard cells is typically performed after the macro placement. However, an important consideration is that standard cells are often placed in clusters based on their logic hierarchy. If a cluster is large enough, it can be modeled as a ``macro''. Therefore, the placement of standard cell clusters in turn impact how macros are placed. This is reflected in the current design process, where physical design teams typically rely on feedback from front-end design teams to create guide or fence regions for standard cell clusters, which are crucial for floorplanning and constraint-based cell placement. The impact of standard cells on macro placement cannot be neglected. This is illustrated in Fig. \ref{fig:dataflow illustration}, where routing resources can be optimized by adjusting the location of \textit{Macro D}, considering macro-cell connections and dataflow among these modules. Additionally, the orientation of the macro plays a key role in determining overall QoR, as different numbers of pin accesses are available on each side of the macro, leading to different wirelengths. This is also shown in Fig. \ref{fig:dataflow illustration}, where \textit{Macro D} is flipped along the Y-axis to reduce the distance to the logically connected cell clusters. Inspired by these observations, we propose DAS-MP, which incorporates a series of incremental optimization strategies for high-quality macro placement. First, recognizing the importance of dataflow awareness, we introduce a novel methodology that efficiently extracts four distinct types of ``hidden'' relationships between macros and cell clusters. To further refine the macro placement after the initial optimization, we integrate macro area consideration and orientation optimization into the fine-tuning process. In this process, macros of varying sizes are adjusted based on dataflow intensity, and macro orientation is optimized to align with the dataflow direction. By refining the placement in this way, the proposed methodology achieve better correlation between dataflow-aware placement and practical physical design constraints, leading to improved overall performance. Furthermore, the proposed DAS-MP methodology demonstrates excellent portability and efficiency; it can be seamlessly integrated as a plugin with commercial tools, thereby enhancing macro placement outcomes.

The rest of the paper is organized as follows. Section \ref{BackgroundandMotivation} reviews state-of-the-art macro placers and outlines the motivation behind the proposed work. Section \ref{overview} provides an overview of the DAS-MP methodology. Sections \ref{macro-cell-flow} and \ref{cell-cell-flow} detail the extraction algorithms. The fine-tuning methods are discussed in Section \ref{fintuning}. Evaluation results are presented in Section \ref{evaluation-results}. Finally, Section \ref{conclusions} concludes the paper.



\begin{table}[t]
\centering
\caption{A summary of recently proposed macro placers}
\label{tab: related work}
\resizebox{\columnwidth}{!}{
\begin{tabular}{lllcc}
\toprule[1pt]%
                                 \textbf{Type}   & \textbf{Work} & \makecell[l]{\textbf{Key} \\ \textbf{Method}}      & \makecell[l]{\textbf{Dataflow} \\\textbf{Consideration}}  & \makecell[l]{\textbf{Flipping} \\\textbf{Optimization}}     \\ \midrule
\multirow{10}{*}{\makecell[l]{ML-based \\ Methods}}    
                                    & \cite{Mirhoseini2021}   & RL + GNN & \ding{55}    & \ding{55}  \\
                                    & \cite{liu2022floorplanning}  & Graph Attention  & \ding{55} & \ding{55}  \\
                                    & \cite{Cheng2021}        & RL       & \ding{55}   &  \ding{55}    \\
                                    & \cite{Vashisht2020}     & RL + SA     &   \ding{55} &    \ding{55}        \\
                                    & \cite{agnesina2023autodmp} & NN + Bayesian & \ding{55} & \ding{55}  \\ 
                                    & \cite{Lin2021}          & NN    &  \ding{55}     &  \ding{55}      \\ 
                                    & \cite{oh2022bayesian} & Bayesian&\ding{55} & \ding{55} \\
                                    & \cite{Ward2012}         & SVM + NN   & \checkmark  &   \ding{55}         \\
                                    & \cite{Cheng2018}        & Training-based Prediction & \ding{55}  &  \ding{55}      \\
                                    & \cite{Liu2021}          & Gradient Optimization  & \ding{55} &  \ding{55}    \\
                                    
                                    \midrule
\multirow{13}{*}{\makecell[l]{Traditional \\Methods}} 
                                    &\cite{alpert2005semi}    & Priority-queue clustering & \ding{55} &  \ding{55}   \\
                                    & \cite{Pan2006}          & Linear Programming      & \ding{55} &  \ding{55}  \\
                                    &\cite{chen2007mp}        & MP-tree & \ding{55}  &  \ding{55}   \\
                                    & \cite{Lui}              & Electrostatics-based  & \ding{55} &  \ding{55}     \\
                                    & \cite{Cheng2019}        & Density Function  & \ding{55} & \ding{55}      \\
                                    & \cite{Chen2008}         & Linear Programming    &\ding{55} &  \ding{55}   \\
                                    &  \cite{chen2014routability, chiou2016circular}        & CP-tree & \ding{55} &  \ding{55}   \\
                                    &\cite{hsu2013routability}& Hierarchical Approach& \ding{55} & \ding{55}  \\
                                    & \cite{Ye2000}           & Dataflow Graph & \checkmark  & \ding{55}         \\  
                                    & \cite{Vidal-Obiols2019,Vidal-Obiols2021}
                                    & Dataflow Graph      &  \checkmark  &  \ding{55}  \\
                                    & \cite{Kahng2022}        & Simulating Human& \checkmark  & \checkmark  \\  
                                    & \cite{kahng2023hierrtlmp} & Hierarchy Clustering & \checkmark  &  \checkmark \\ 
                                    & \cite{10546560} & Dataflow Awareness & \checkmark  &  \ding{55}  \\      
                                \bottomrule[1pt]
                                    
\end{tabular}
}
\vspace{-10pt}
\end{table}

\section{Background and Related Work}
\label{BackgroundandMotivation}

\subsection{Related Work}
\label{relatedwork}
Macro placement and floorplanning have long been fundamental challenges in the field of electronic design automation (EDA). Traditionally, macro placement has been modeled as an objective-driven problem and tackled using analytical methods or iterative-perturbative techniques. The recent rise of machine learning (ML) has spurred interest in applying these techniques to macro placement, leading to the development of reinforcement learning (RL) and deep learning (DL)-based methods. As summarized in Table \ref{tab: related work}, we categorize these techniques into two groups: ML-based methods and traditional methods, which refer to non-ML-based approaches.


\textbf{For ML-based methods}, Google \cite{Mirhoseini2021} utilized Graph Neural Network (GNN) to process netlists and applied Reinforcement Learning to constrain the placement of macro blocks while considering some parameter constraints. This approach introduced a new dimension for macro placement and has generated significant interest in the field. In \cite{Cheng2021}, building on Google's method, a Convolutional Neural Network (CNN) was used to capture global information, and a GNN was used to extract more detailed information, and then fully connected into policy to achieve multi-perspective optimization. Similarly, \cite{liu2022floorplanning} presented a graph attention-based floorplanner to learn an optimized mapping between circuit connection and physical wirelength. While in \cite{Vashisht2020}, a combined approach using simulated annealing and RL was employed for macro placement. 
In \cite{agnesina2023autodmp}, AutoDMP was proposed as an optimization framework built on DREAMPlace \cite{Lin2021}, incorporating multi-objective Bayesian optimization into the design space search. It demonstrated the ability to further improve the results obtained from DREAMPlace. Similarly, \cite{oh2022bayesian} modeled the macro placement problem as Bayesian optimization over sequence pairs, concluding that Bayesian optimization is more sample-efficient than RL and can accommodate more realistic objectives. In \cite{Ward2012}, dataflow was modeled as a graph and solved through iterative optimization using Support Vector Machines (SVM) and Neural Networks (NN). In \cite{Cheng2018}, a machine learning-based method was introduced to predict half-perimeter wirelength (HPWL) and routing congestion directly after macro placement. \par

\textbf{For traditional methods}, macro placement has been viewed as an object-driven optimization problem. In \cite{alpert2005semi}, a priority-queue data structure was used to repeatedly cluster the globally best pair of objects. \cite{Pan2006} proposed a linear programming method, combining the grid-based method and the path-based method, by considering the critical path of timing disturbances during the placement optimization process. \cite{chen2007mp} proposed a multi-packing tree (MP-tree) representation to handle mixed-size designs, which can effectively optimize macro locations and facilitate standard cell placement and routing. In ePlace \cite{Lui}, the Nesterov method was used as the nonlinear solver, and the Lipschitz constant dynamically predicted the step size. While RePlace \cite{Cheng2019} mainly optimized density and presented a method for adapting density penalties through improved dynamic step size adaptation. \cite{Chen2008} proposed a new adaptive SA process that uses smoothing techniques for cost evaluation to reduce the number of linear program solutions. \cite{chen2014routability, chiou2016circular} constructed circular-packing trees (CP-trees) that can circularly position macros along the chip boundaries. In \cite{hsu2013routability}, a hierarchical approach was proposed to accelerate turnaround time. \par

Dataflow has been one of the key features used by engineers to optimize macro positioning.
In \cite{Ward2012}, the dataflow was modeled as a graph and solved through SVM and NN iterative optimization. In \cite{Ye2000}, optimized data path layout was generated by utilizing rule information directly extracted from dataflow graphs. \cite{lin2021dataflow} also leverage dataflow to address the problem of preplaced macros, which cannot be handled by simulated annealing (SA). A recently released macro placer named RTL-MP \cite{Kahng2022} considered the macro-to-macro dataflow and proved that such dataflow is helpful to get a human-quality layout. An updated version of this placer was presented in \cite{kahng2023hierrtlmp}. These two method also considerting macro flipping for further macro position optimization. Recently, commercial tool vendors also started to pay attention to dataflow analysis and equip some of their tools with such capability \cite{maxeda,synopsysplacer,cadenceplacer}. However, among these dataflow-aware macro placers, only a few have considered the impact of macro-to-cell or cell-to-cell dataflow on macro placement. Many approaches have failed to effectively leverage this information to constrain subsequent steps, resulting in little to no improvement—or even a decline—in design quality. In \cite{10546560}, a methodology was proposed to extract various dataflow connections and incorporate this information into macro placement using simulated annealing (SA). While this work demonstrated the effectiveness of dataflow, it treated all macros equally without considering differences in macro area and failed to account for macro orientation, which is critical for QoR optimization.


\subsection{Design Principles and Contributions}
The proposed work differs from the existing macro placers in the following design principles. The proposed methodology involves four different types of connections including cell connections and placement in the macro placement process and defines a new dataflow-aware optimization target. The proposed dataflow extraction methodology offers a good starting point for any macro placers or mixed-size placers such as \cite{7008518}. It complements the existing tools with newly discovered ``hidden'' connections that are essential for the final QoR. The key contributions from DAS-MP are summarized as follows.
\begin{itemize}
    \item \textbf{Introducing New Cell Dataflow Relationships:} Unlike previous work that primarily focuses on macro-to-macro dataflow, the proposed work recognizes that cells occupy substantial area, and their placement significantly influences macro positioning.
    Therefore, we propose to extract several new dataflow relationships that take into account for cell positions. This intuition is illustrated in Fig. \ref{fig:dataflow illustration}, where routing resources are optimized by adjusting the location of \textit{Macro D}, considering macro-cell connections and the flow of data among these modules.    
\item \textbf{Expanding Macro-Macro Dataflow:} Beyond macro-cell relationships, we extend our approach by introducing multi-hop connections, such as macro-cell-cell interactions. Additionally, we incorporate indirect macro-macro relationships to capture more complex dependencies between macros. This comprehensive dataflow model enhances the optimization process during macro placement.
    \item \textbf{Introducing Two-step Incremental Fine-tuning:} Fine-tuning is essential to further refine the macro placement after the initial optimization, addressing local inefficiencies that may not be captured during global optimization. The fine-tuning process is divided into two key stages. 
    \begin{itemize}
        \item First, we introduce a feedback mechanism to adjust the weights of macro connections. This mechanism refines placement by accounting for the effect of macro area on congestion. Larger macros, which tend to increase congestion, are strategically placed to minimize their impact, while smaller macros are positioned more flexibly to improve overall placement quality. This feedback mechanism ensures that placement optimization considers both connectivity and macro size, enhancing layout efficiency.
        \item Second, we optimize the orientation of macros to reduce wirelength and improve placement efficiency. Directional flipping is considered for each macro, with the flipping order determined based on topological relationships in the dataflow, focusing on minimizing congestion and aligning macros with the dataflow direction.
    \end{itemize}
\item \textbf{Demonstrated Placement Outcome:} Extensive experiments on diverse benchmarks demonstrate significant improvements: an average improvement of 7.9\% in HPWL compared to a state-of-the-art dataflow-aware macro placer RTL-MP \cite{Kahng2022}, with a maximum improvement of 12.1\%. Congestion overflow is reduced by 82.5\% on average, with a runtime overhead of less than 1.5\% of the total macro placement time. As a result, these improvements translate to approximately 36.97\% and 59.44\% improvement on WNS and TNS compared to RTL-MP \cite{Kahng2022}.

\end{itemize}

\section{DAS-MP Methodology Overview}
\label{overview}
\begin{figure}[!t]
    \centering
    \includegraphics[width =0.98\linewidth]{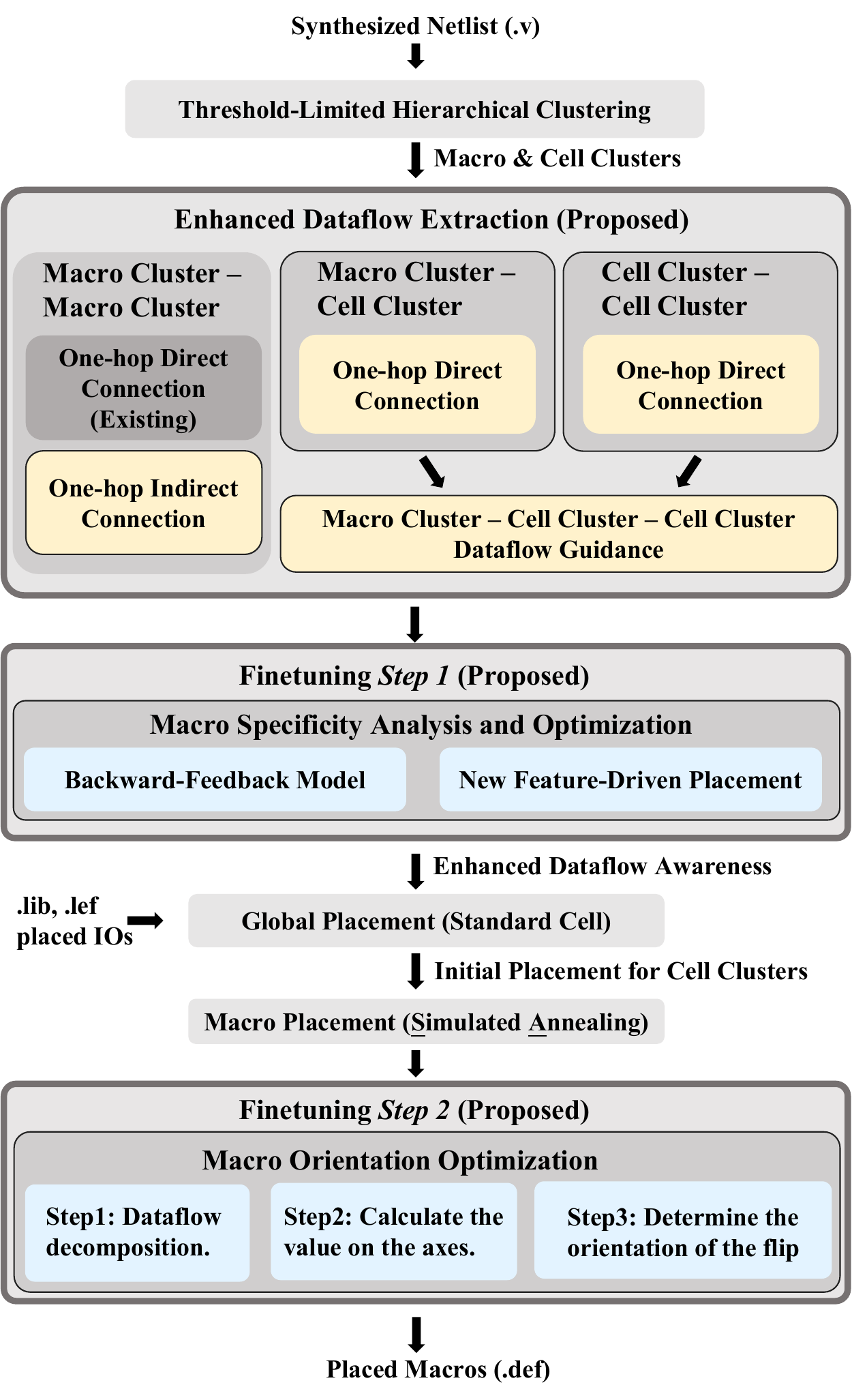}
    \vspace{-10pt}
    \caption{The overview of the overall DAS-MP methodology. The first part is the proposed enhanced dataflow extraction process where light yellow rectangles outline all the newly proposed features in dataflow extraction. The second and third parts are the proposed fine-tuning stages where light blue rectangles list the macro specificity analysis and orientation optimization.}
    \label{fig:overview}
    \vspace{-5pt}
\end{figure}

The proposed DAS-MP methodology and fine-tuning process, which incorporates the proposed dataflow connection extraction, is depicted in Fig. \ref{fig:overview}, with the key steps explained below.\par
The first step is \textbf{threshold-limited hierarchical clustering}, where the synthesized netlist is fed into a clustering engine adapted from a recently proposed hierarchical auto-clustering method in \cite{Kahng2022}.
This engine converts the structural netlist representation of the RTL design into a clustered netlist. Clusters are split and merged when the number of macro or standard cells exceeds or falls below a preset threshold. Bundled I/O pins are modeled as macro clusters in this process. Clustering also reduces the search cost for later connection extraction steps. Each cluster contains either macros or standard cells, and a cross-cluster connection is defined as a \textbf{one-hop connection}.\par
With the generated macro or cell clusters, we explore the \textbf{dataflow extraction} in three categories, as shown in the top box of Fig. \ref{fig:overview}. The first category includes connections among macro clusters, where most prior work focused on the one-hop direct connections between macros. We enhance this connection extraction proces by also considering potential indirect connections between macro clusters resulting from shared connections with the same cell clusters. The second category of dataflow connections captures dataflow between macro clusters and cell clusters, where logical connections and their strengths are extracted and stored for later use in placement optimization. The third category involves connections among cell clusters, which were typically overlooked in previous macro placers but still indirectly affect macro placement and are therefore incorporated into our model. The connections between {\tt macro cluster-cell cluster} and {\tt cell cluster-cell cluster} are combined into a unified {\tt macro cluster-cell cluster-cell cluster} dataflow guidance to further improve macro placement.\par

With the extracted dataflow, a fine-tuning step (denoted as ``\textbf{Finetuning Step 1}'' in the figure) refines the dataflow by incorporating the area characteristics of each macro. For a 2-hop dataflow connection, represented as {\tt macro cluster-cell cluster-cell cluster}, we propose a weight feedback model. In this model, the weight of the second-hop cell cluster is directly applied to the macro requiring optimization. Additionally, a new feature representing macro area is introduced to account for the impact of different macro sizes on the final placement. 

After the first fine-tuning process, the enhanced dataflow awareness provides a comprehensive view of data movement, considering macro uniqueness within the design. Since macro placement depends heavily on cell clusters, we first conduct a global placement (GP) to position the cell clusters. Macros are then placed by incorporating the extracted dataflow information into a loss function, which is optimized using the Simulated Annealing (SA) algorithm \cite{kirkpatrick1983optimization, Kahng2022} and represented in the netlist using Sequence Pair \cite{murata1996vlsi}.  
Depending on the proximity to the convergence point, GP and macro placement can be run iteratively.\par  
Once the dataflow-guided placement is complete, a second fine-tuning process (denoted as ``\textbf{Finetuning Step 2}'' in the figure) is performed to refine macro orientation, which significantly impacts placement quality. Our method’s extracted dataflow enables the determination of both dataflow direction and weight. This process involves three steps: first, weighted dataflow decomposition; second, calculation of values along the axes; and third, orientation adjustment for each macro. Once this fine-tuning is complete, the macro placement from DAS-MP is finalized, allowing subsequent physical design stages to proceed.

\begin{figure}[t]
    \centering
    \subfigure[One-hop direct dataflow connection of macro cluster-cell cluster.]{
    \includegraphics[width = 0.9 \linewidth]{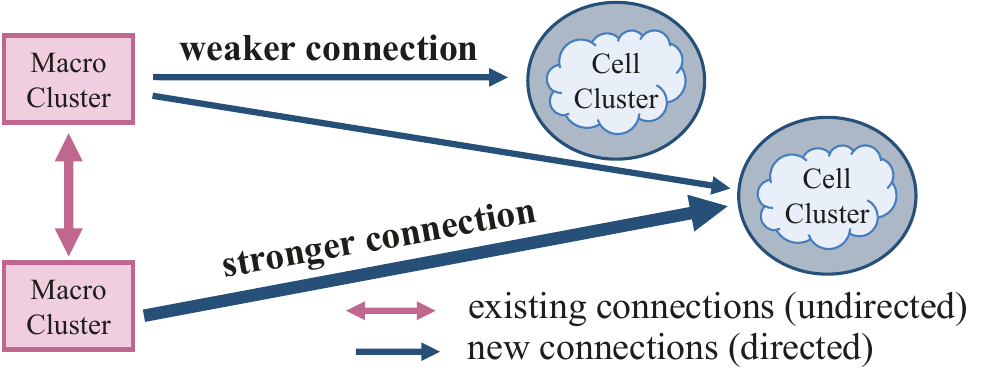}
    \label{fig:m-c}
    }
    \subfigure[one-hop indirect dataflow connection from macro cluster-macro cluster.]{
    \includegraphics[width = 0.9 \linewidth]{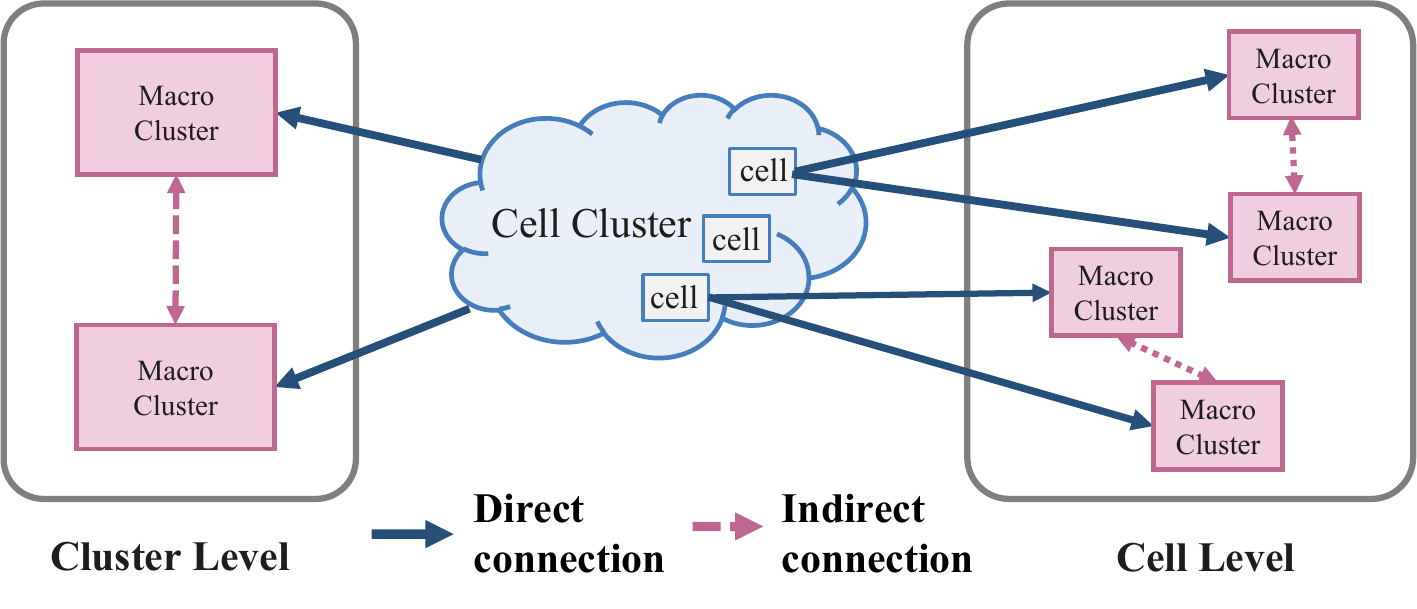}
    \label{fig:indi m-m}
    }
    \caption{One-hop direct dataflow connection of macro cluster-cell cluster. The strength is given based on the bit width of the extracted connections. Illustration of one-hop indirect dataflow connection from macro cluster-macro cluster. The left figure treats the cell cluster as the smallest unit, and the right figure treats the single cell instance as the smallest unit.}
\end{figure}

\section{Dataflow Between Macro Clusters and Cell Clusters}
\label{macro-cell-flow}
Existing macro placers primarily focus on direct connections among macros. However, given the vast number of cell instances in the design, analyzing the logical relationships between macro clusters and cell clusters is essential. The {\tt macro cluster-cell cluster} connections can reveal indirect and virtual links between macros. This section details the connection extraction process.

\subsection{One-Hop Direct Macro Cluster-Cell Cluster Dataflow}
\label{onehopmacro-cell}
The dataflow from the netlist is converted into a graph, where the prior work mostly modeled the {\tt macro cluster-macro cluster} connections as undirected. As can be seen in Fig. \ref{fig:m-c}, in addition to the undirected {\tt macro cluster-macro cluster} connections, we also consider directionality when extracting connections between macros and cells. This approach enables the identification of more types of logical relationships. Although brute force search (BFS) is used, we introduce pruning in the search process and optimize data storage during graph building and traversal. The strength of connections is defined based on the data bit-width. 
To incorporate the impact of the dataflow, a loss function is defined in Equation \ref{eq:loss 1}. This loss function is utilized to guide the subsequent solution search process, which is iterated to optimize the wirelenghth as the final quality indicator. In Equation \ref{eq:loss 1}, $WL$ is half perimeter wire length (HPWL) and $w_x$ is a weight factor which is defined by dataflow $bit\_width$ between the clusters.

\begin{gather}
   w_{0}=bit\_width(macro\_cluster_1,macro\_cluster_2)\notag \\
    w_{1}=bit\_width(macro\_cluster,cell\_cluster)\notag \\
    loss = w_{0} * WL_{macro-macro} + w_{1} * WL_{macro-cell}
    \label{eq:loss 1}
\end{gather}

\begin{algorithm}[t]
    \footnotesize
    \caption{Indirect Macro Dataflow Connection Extraction Algorithm}
    \label{alg:indirect algorithm}
    \begin{algorithmic}[1]
        \Require $.lef$, $.lib$, $.v$, $constraints$  
        \Ensure two types of indirect macro connections  

        \State \textcolor{blue}{\Comment{At cell cluster level (Fig. \ref{fig:indi m-m} left)}}
        \For{each cell cluster (in cell-macro connection)}  
            \State $vector:cell\_connected\_macro\_vec$
            \State $num = cell\_connected\_macro\_vec.size$
            \For{i = 0 to num }  
                \For{j = i to num}
                    \State $macro\_src = cell\_connected\_macro\_vec[i]$ 
                    \State $macro\_sink = cell\_connected\_macro\_vec[j]$ 
                    \State $addconnection\ (macro\_src, macro\_sink)$  
                \EndFor
            \EndFor
        \EndFor
        \State  \textcolor{blue}{\Comment{At cell instance level (Fig. \ref{fig:indi m-m} right)}}
        \For{each vertex in cell fanin map} 
            \State $vector:same\_vertex\_macro\_vec$
            \For{each $pin$ (in vertex pin list)}
                 \State $id$ =
                 find the $macro\_id$ where $pin$ located
                \If{$id$ is in a macro cluster}
                    \State $same\_vertex\_macro\_vec.push\_back(id)$ 
                \EndIf
            \EndFor      \textcolor{blue}{\Comment{Find macros connected to the same cell}}
            \State $total = same\_vertex\_macro\_vec.size$
            \For{i = 0 to $total-1$}
                \For{j = i+1 to total}
                    \State $macro\_src = same\_vertex\_macro\_vec[i]$
                    \State $macro\_sink = same\_vertex\_macro\_vec[j]$ 
                    \State $addconnection\ (macro\_src, macro\_sink)$ 
                \EndFor
            \EndFor
        \EndFor \\
        \Return indirect macro connections
    \end{algorithmic}
    \vspace{-2pt}
\end{algorithm}

\subsection{One-Hop Indirect Macro-Macro Dataflow}

For certain designs, macro clusters may not have a direct connection among themselves, but they may share connections with the same cell clusters. Such connections can indicate indirect dataflow among macro clusters. Then these ``hidden'' relations can be converted into virtual connections to guide macro placement. This type of connection is challenging to find due to the scale of the search and the complexity involved in excluding common signals that are shared among all macros, such as clock and reset signals. In the proposed methodology, we extract these virtual connections in a hierarchical way. This has been shown in Fig. \ref{fig:indi m-m}. 

Firstly, we treat the \textit{cell cluster} as the smallest unit and examine all macro connections to or from the same cell cluster (shown on the left part of Fig. \ref{fig:indi m-m}). We view these macros as being virtually connected and convert these connections into virtual dataflow among macro clusters.
This process has been shown in lines 2-12 of Algorithm \ref{alg:indirect algorithm}. After traversing all the clusters, we establish virtual connections between macros that share common connections to the same cell cluster (shown Fig. \ref{fig:indi m-m} left).
Direct connections from a cell cluster to multiple macro clusters are then transformed into indirect connections between macro clusters within the cluster-based graph. In this graph, vertices represent the mapping nodes of fanout and pin, and the $addconnection$ function creates the required virtual connections.

In the second part of the extraction process, we treat the \textit{single standard cell} instance as the smallest unit and check their macro fanouts. If a single cell drives multiple macros, we view these macros as being virtually connected (Fig. \ref{fig:indi m-m} right). Line 14-21 in Algorithm \ref{alg:indirect algorithm} details the extraction process of such connections. By far, we have supplemented the relationship among macro clusters by considering all direct and indirect connections between macro clusters. It is worth mentioning that in this example, we traverse all clusters to find these virtual connections. However, one can also selectively extract only cluster-level connections to reduce the search complexity.

We also define the new loss function for the indirect macro cluster connection in Equation \ref{eq:loss 2}, where $w_i$ represents the sum of dataflow $bit\_width$ between two macro clusters to cell cluster and $WL$ denotes the HPWL.

\begin{equation}
\label{m-m weight}
    \begin{split}
                w_i = &bit\_width(macro\_cluster_1,cell\_cluster) \\
             +&bit\_width(macro\_cluster_2,cell\_cluster) \\
    \end{split}
\end{equation}
\vspace{-10pt}
\begin{gather}
\label{eq:loss 2}
loss_{indirect\_macro} = w_i * WL_{indirect\ macro-macro}     
\end{gather}

\section{Dataflow Among Cell Clusters}
\label{cell-cell-flow}

\begin{figure}[t]
    \centering
    \includegraphics[width = 0.9 \linewidth]{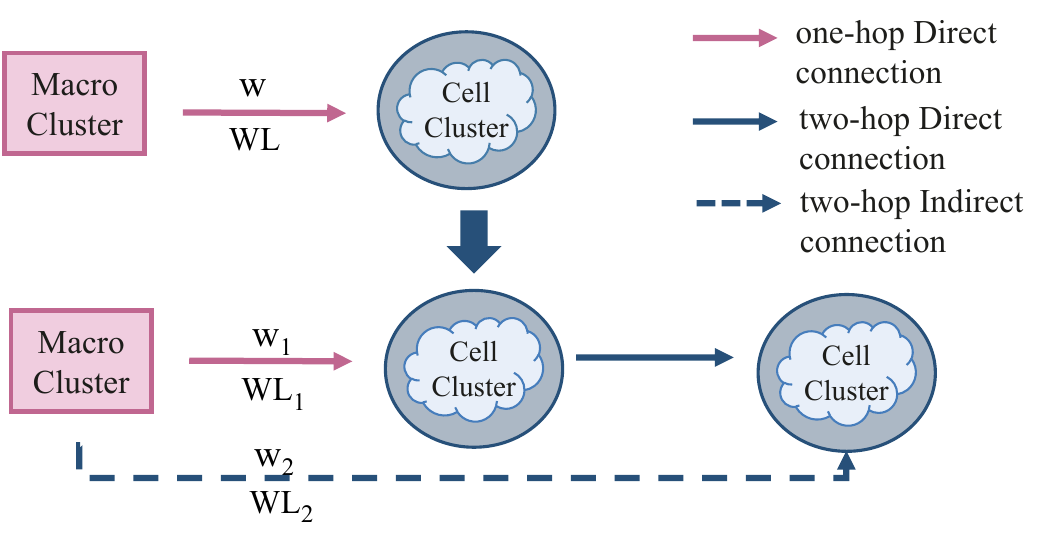}
    \vspace{-10pt}
    \caption{Illustration of newly established two-hop dataflow connection macro cluster-cell cluster-cell cluster.}
    \label{fig:m-c-c}
\end{figure}
\begin{figure}[t]
    \centering
    \includegraphics[width=0.95\linewidth]{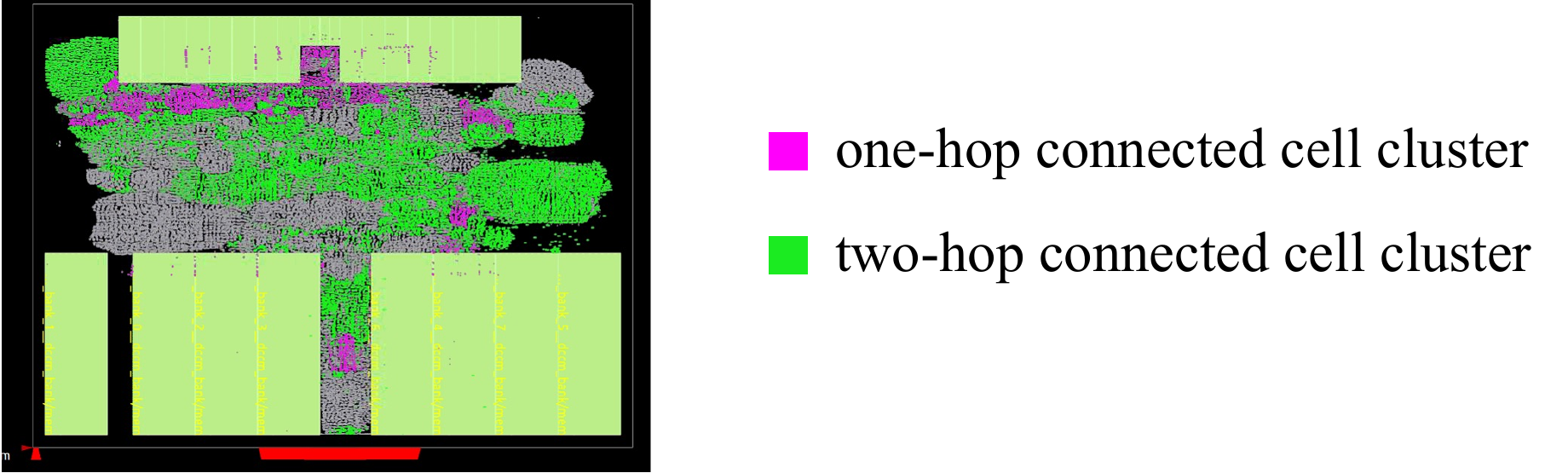}
    \vspace{-5pt}
    \caption{Comparison between one-hop connected cell cluster (only macro cluster-cell cluster) and two-hop connected cell cluster (after considering macro cluster-cell cluster-cell cluster) in an example design {\tt swerv\_wrapper}. }
    \label{fig:fig6}
\end{figure}

\begin{algorithm}[t]
    \footnotesize
    \caption{Macro Cluster-Cell Cluster-Cell Cluster Connection Extraction Algorithm (Fig. \ref{fig:m-c-c})}
    \label{alg:greedy algorithm}
    \begin{algorithmic}[1]
        \Require $.lef$, $.lib$, $.v$, $constraints$  
        \Ensure connections of macro cluster-cell cluster-cell cluster  
        \State $map: macro\_cell\_connect$ 
        \For{each connection} 
            \If{src is macro \&\& sink is cell}    \textcolor{blue}{\Comment{First connect macro-cell}}
                \State $addconnection\ (macro\_src, cell\_sink)$
                \State $macro\_cell\_connect[sink.id].first = 1$
                \State $macro\_cell\_connect[sink.id].second = src\_id$
            \EndIf
            \If{src is cell \&\& sink is cell}        \textcolor{blue}{\Comment{Then connect cell-cell}}
                \If{macro\_cell\_connect[src\_id].first == 1} \\\textcolor{blue}{\Comment{Checking whether this cell is connected to a macro}}
                    \If{src\_id != sink.first}
                        \State $macro\_id = macro\_cell\_connect[src\_id].second$ 
                        \State $addconnection\ (macro\_id, cell\_sink)$ 
                        \State $addconnection\ (cell\_src, cell\_sink)$ 
                    \EndIf
                \EndIf
            \EndIf
        \EndFor \\
        \Return macro cluster-cell cluster-cell cluster connections
    \end{algorithmic}
\end{algorithm}

The ultimate goal of this work is to establish  {\tt macro cluster-cell cluster-cell cluster} dataflow to further analyze the impact of multi-hop dataflow connection.
This dataflow consists of two parts: the {\tt macro cluster-cell cluster} connection discussed in Section \ref{onehopmacro-cell} and the {\tt cell cluster-cell cluster} connection. The size of cell clusters is also considered, as larger clusters typically have a greater impact on the positioning of virtually connected macros. Therefore, the connection between cell clusters is weighted using the product of a constant $k$, $bit\_ width$, area, and instance number, as shown in Equation \ref{weight-function}.
\begin{equation}
w_{j} = k * bit\_width * cluster_{area} * cluster_{number}
\label{weight-function}  
\end{equation}
 
After extracting the dataflow connections among cell clusters using Algorithm \ref{alg:greedy algorithm}, we convert them into two-hop connections between macro clusters and cell clusters, as shown in Fig. \ref{fig:m-c-c}. The conversion assigns higher weights to two-hop virtual connections compared to one-hop direct connections. In lines 3–7, we first label the macro clusters directly connected to the cell cluster. After traversing the connections between cell clusters, we check for any labeled macro clusters. Once identified, a {\tt macro cluster-cell cluster-cell cluster} dataflow connection is established. Fig. \ref{fig:fig6} illustrates the impact of a two-hop connection using the {\tt swerv\_wrapper} design, showing that green clusters with second-hop connections to macros occupy a much larger area and will impact macro positioning. The final updated definition of the loss function is shown in Equation \ref{equation_final_loss}, 
\begin{gather}
     loss = w_0 * WL_{m-m} + w_1 * WL_{m-c} + w_2 * WL_{m-c-c}
    \label{equation_final_loss}   
\vspace{-15pt}    
\end{gather}
where $w_0$, $w_1$ and $w_2$ are defined in Equation \ref{m-m weight}, Equation \ref{eq:loss 1} and Equation \ref{weight-function}, and $WL_{m-m}$ refers to total HPWL of {\tt macro cluster-macro cluster} connections, $WL_{m-c}$ means HPWL of {\tt macro cluster-cell cluster} connections, $WL_{m-c-c}$ represents HPWL of the second-hop connection from {\tt macro cluster-cell cluster-cell cluster}. 

\section{Fine-tuning Processes based on Dataflow}
\label{fintuning}
\subsection{Macro Specificity Analysis and Optimization}
\begin{figure}[t]
    \centering
    \subfigure[2-hop backward-feedback weight update.]{
    \includegraphics[width=0.85\linewidth]{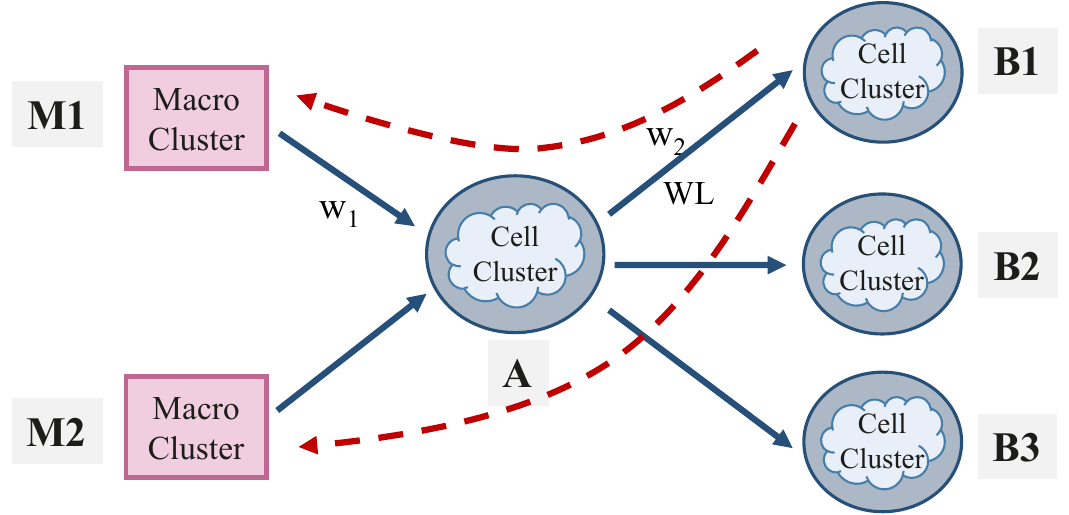}
    }
    \subfigure[2-hop area-driven placement optimization.]{
    \includegraphics[width=0.85\linewidth]{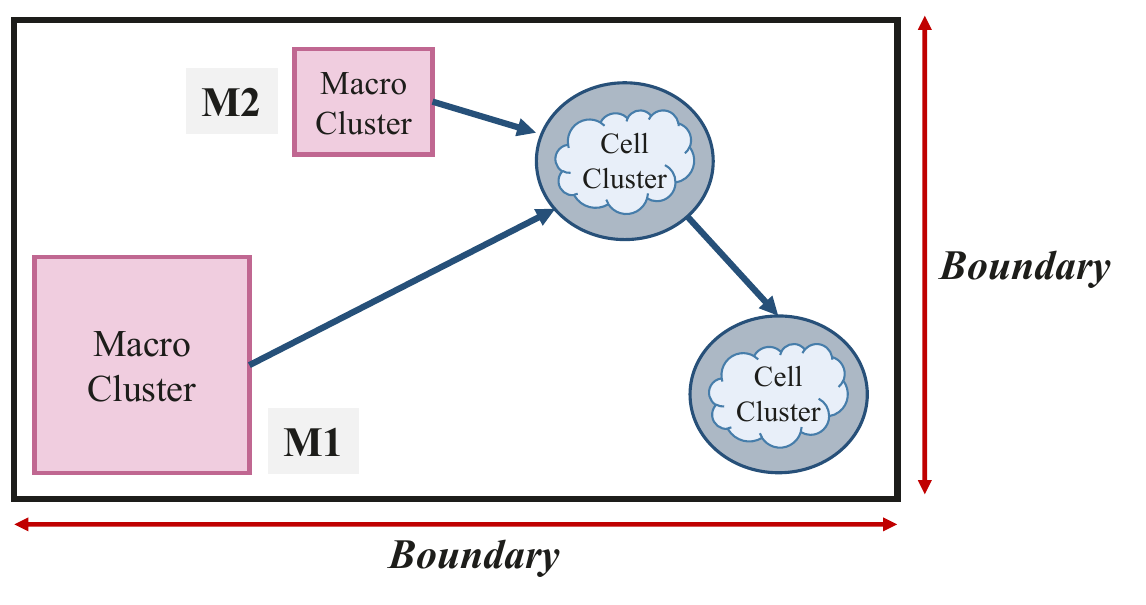}
    }
    \caption{Fine-tuning for Macro Specificity Analysis and Optimization.}
    \label{fig:finetuning1}
\end{figure}

So far all macros were treated equally when extracting dataflow relationships, without considering their individual characteristics, especially area. Larger macros typically occupy more chip area and are more prone to causing congestion or other placement violations. To address this, we propose a feedback model that incorporates macro characteristics into the loss function during the Simulated Annealing process, refining connectivity and placement based on macro size. Specifically, the model adjusts the weights of two-hop connections in a cascading manner, where the influence of second-hop cell clusters is weighted differently based on their connection to source macros. Additionally, macro area within the cluster is factored into the placement optimization, as larger macros are more likely to impact floorplan congestion, while smaller ones allow greater placement flexibility. The final placement is further refined based on these weighted adjustments, producing an optimized floorplan. While other features, such as the number of cell clusters, could also influence placement, they are omitted here to maintain variable uniformity. The following sections detail the methodology and implementation.

\subsubsection{\textbf{Macro Characteristic Backward-feedback Model}} 
As discussed in Section \ref{cell-cell-flow}, the 2-hop {\tt macro-cell-cell} connectivity relationship is reflected by $w_2 * WL_{m-c-c}$ in the loss function in Equation \ref{equation_final_loss}. Here, only the strength of the connectivity between cell clusters was considered. While this approach links macros to two-hop cell clusters, it only captures the connection strength between one-hop and two-hop cell clusters without distinguishing the unique influence of different source macros. As shown in Fig. \ref{fig:finetuning1}(a), the connection strength between the B1 cell cluster and the two macros (M1 and M2) remains identical, despite the differences in the source macros’ characteristics. Since the goal is to improve macro placement, it is essential to account for the varying connectivity strength between the same two-hop cell cluster and different macros. To address this, we propose coupling the weights of one-hop macro-cell clusters with the weights of two-hop cell-cell clusters. This allows the feedback from the source macro’s characteristics to propagate through the connectivity network, creating differentiated connection strengths for different source macros, as shown in Equation \ref{finetuning_0}. This coupling reflects both the local connection strength and the global impact of macro characteristics, enhancing the accuracy of the placement optimization process.
\begin{gather}
    loss_{m-c-c} = w_1 * w_2 * WL_{m-c-c}
     \label{finetuning_0}   
\end{gather}

\subsubsection{\textbf{Feature-driven Placement Optimization}} 
Beyond multi-hop connection strength, macro area significantly impacts floorplan congestion and overall placement quality. Based on back-end design experience, larger macros should be placed near the chip boundary to reserve more space for cell clusters, while smaller macros should be positioned closer to cell clusters to enhance placement performance. To integrate this consideration, macro sizes are incorporated into the final loss function. As illustrated in Fig. \ref{fig:finetuning1}(b), the larger macro cluster M1 is strategically positioned at the chip boundary, while the smaller macro cluster M2 is placed closer to the cell cluster, influenced by the cell cluster’s pull. This area-based placement strategy ensures optimal utilization of available space and reduces routing congestion, thereby improving overall design efficiency.


\subsubsection{\textbf{New Loss Function for Optimizing the Macro Placement}}
Based on the fine-tuning processes described in the prior sections, we introduce a refined loss function that incorporates both the connection weights of 2-hop macro cluster-cell cluster-cell cluster connections and macro area into the simulated annealing process. In Equation \ref{area}, the macro area $A_i$ is normalized to the interval [1,2] based on the maximum and minimum macro areas in the design, yielding $A_i'$. The loss function for the 2-hop connection is then updated by adjusting the weights and incorporating the area constraints, as shown in Equation \ref{finetuning}.
\begin{gather}
A_i' = 1 + \frac{(A_i - A_{\min})}{A_{\max} - A_{\min}} 
    \label{area}   
\vspace{-5pt}    
\end{gather}

\begin{gather}
    loss_{m-c-c} = \frac{\sqrt{w_1 * w_2}}{A_i'} * WL_{m-c-c}
     \label{finetuning}   
\end{gather}

In Equation \ref{area}, the macro area $A_i$ is normalized to the interval [1,2] based on the maximum and minimum macro areas in the design, resulting in $A_i'$. The loss function for the 2-hop connection is then updated by incorporating the revised weights and area constraints, as shown in Equation \ref{finetuning}.
By leveraging the final SA loss function, we obtain SA-based macro placement results, which serve as the foundation for further macro orientation optimization through flipping.

\subsection{Macro Orientation Optimization}
Most automatic floorplanning tools focus on macro placement, often neglecting macro flipping, which can effectively reduce wirelength and improve timing and other design metrics \cite{incredflip}. In advanced technology nodes, macro orientation is typically restricted to 0 and 180 degrees to prevent layer misalignment. If rotation is necessary, it is limited to flipping along the x and y axes, effectively capping adjustments at 180 degrees. We define four flipping modes: no flipping (N), x-axis flipping (FN), y-axis flipping (FS), and both x and y-axis flipping (S).

In the fine-tuning process, we begin from a selected point based on the dataflow connection to the I/O pins. The flipping order of subsequent macros is determined by their topological relationships within the dataflow while respecting dataflow-based order constraints. To model the macro flipping problem accurately, we perform dataflow vectorization and decomposition, using superimposed vectors to guide the flipping direction.


Since dataflow inherently encodes the connection information of a design through its directional and weighted nature, making it well-suited for vector-based representation. By mapping dataflow into a vector space ($V_T$), we identify three distinct vector types relevant to macro flipping:
\begin{itemize}
    \item $V_{mm}$ — vectors between macros
    \item $V_{mc}$ — vectors from a macro to cell clusters
    \item $V_{mcc}$ — vectors from a macro to cell clusters at a further hop.
\end{itemize}
As discussed in Section \ref{cell-cell-flow}, we limit the analysis to two hops from macros, as experiments show that standard cells beyond this range have minimal influence on macro positioning. We then analyze the vector projections along the x- and y-dimensions, denoted as $x{V_T}$ and $y{V_T}$, respectively.

\subsubsection{\textbf{Dataflow Decomposition of $V_{mm}$}}
\label{sec:vmm}
\begin{figure}[t]
    \centering
    \subfigure[Macro-Macro dataflow vector ($V_{mm}$) decomposition.]{
    \includegraphics[width=0.8\linewidth]{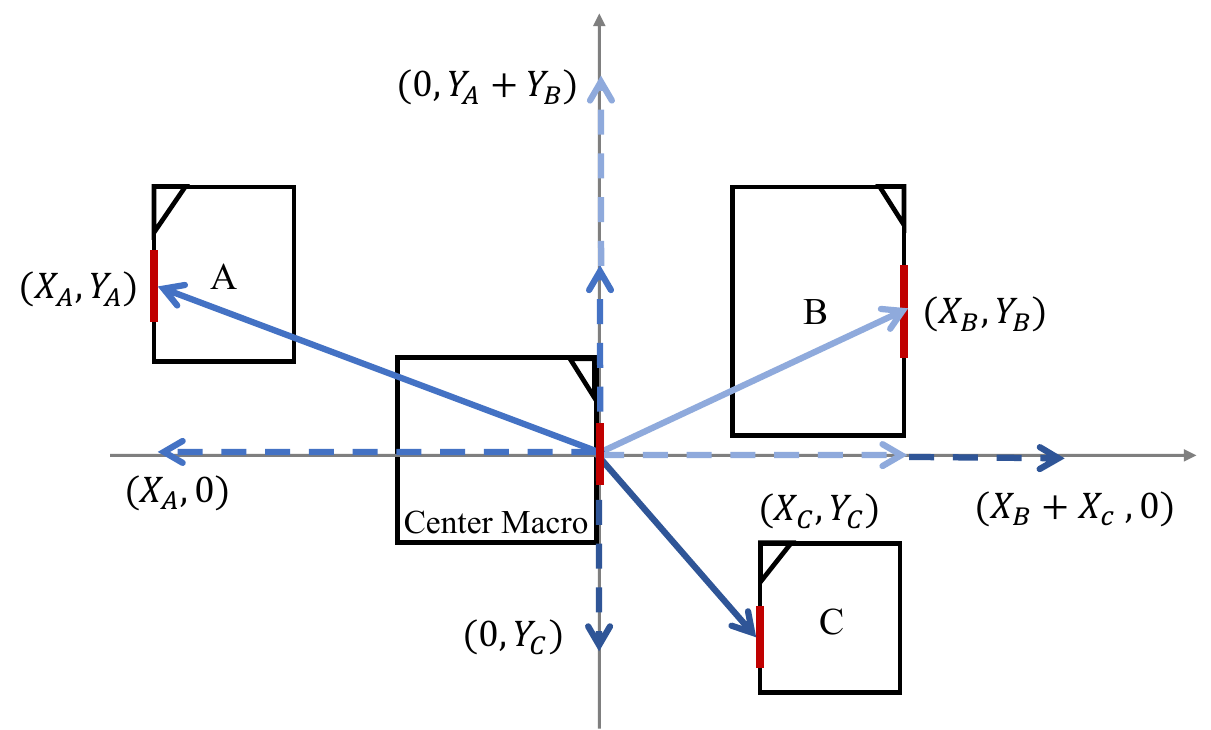}
    \label{fig:flip_mm}
    }
    \subfigure[Macro-Cell ($V_{mc}$) and Macro-Cell-Cell ($V_{mcc}$) dataflow vector decomposition.]{
    \includegraphics[width=0.8\linewidth]{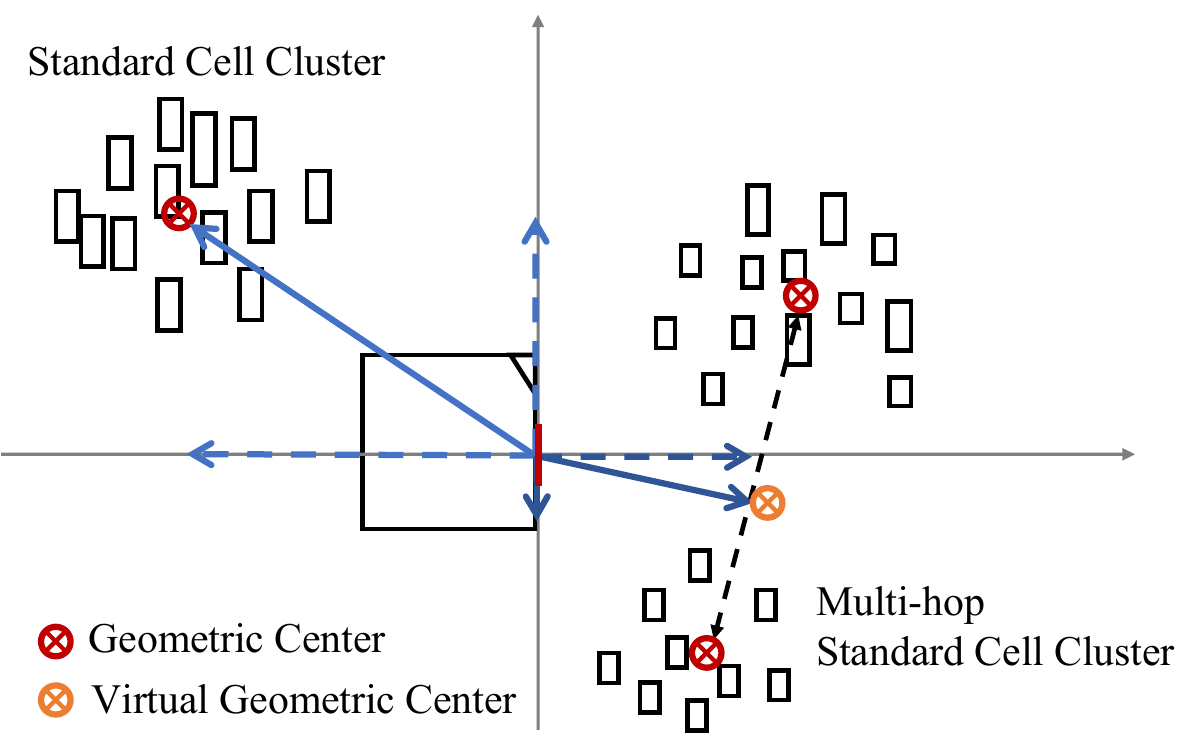}
    \label{fig:flip_mc}
    }
    \caption{Macro orientation optimization guided by enhanced dataflow awareness.}
    \label{fig:macro flipping}
\end{figure}

A key step in decomposing vectorized dataflow is identifying the vector’s starting and ending points. In the proposed method, the origin for decomposition coordinates is set at the midpoint of all pin positions on the target macro. By focusing on pin positions, the macro flipping problem is effectively reduced to accurately positioning these pins within each macro.
In many designs, physical connections between macros are rarely strictly horizontal or vertical and often involve multiple connections, necessitating dataflow decomposition, as shown in Fig. \ref{fig:macro flipping}(a). During decomposition, vertical orientation guides upward or downward flipping, while horizontal orientation guides left or right flipping. The vector direction between macros is determined by the position of the target macro’s pin relative to the connected macro’s pin, with its magnitude calculated as the product of the dataflow exchange’s bitwidth and the physical distance between pin centers (see Equation \ref{eq: define vector}, where $m2$ is the center macro). For macros with multiple out-degree connections, the additive property of vectors allows the overlay of decomposed vectors along the same axis, enabling the calculation of dataflow magnitude in all directions relative to the coordinate axes, as defined in Equation \ref{eq : x + y}, where $n$ is the number of connections for the macro of interest.
\begin{equation}
V_{mm} = bitwidth * ((x_{m_1}, y_{m_1}) - (x_{m_2}, y_{m_2})) 
    \label{eq: define vector}
\end{equation}

\begin{equation}
    \begin{gathered}
        x \{V_{mm}\} = x\{V_{mm_1}\} + x\{V_{mm_2}\} + ... + x\{V_{mm_n}\} \\
        y \{V_{mm}\} = y\{V_{mm_1}\} + y\{V_{mm_2}\} + ... + y\{V_{mm_n}\}
    \end{gathered}
\label{eq : x + y}
\end{equation}

\subsubsection{\textbf{Dataflow Decomposition of $V_{mc}$ and $V_{mcc}$}}
\label{sec:vmc}
The approach to decomposing dataflow between a macro and cells is similar to what has been used for macro-to-macro dataflow.
After obtaining the cell cluster, we estimate its geometric center using Equation \ref{eq: geo center}, as detailed in line 14 of Algorithm \ref{alg:1}, where $c$ is the number of cell instances in a given standard cell cluster, and $x_i$ and $y_i$ are the coordinates of each cell after initial global placement. The dataflow connection between the target macro and the cell cluster is determined by the midpoint of the macro’s pin and the cell cluster’s geometric center, with its weight defined by the dataflow bitwidth. Similar to $V_{mm}$, we project the $V_{mc}$ dataflow vectors onto the x- and y-axes, superimposing these projections as shown in Fig. \ref{fig:macro flipping}(b) and Equation \ref{eq : mc x + y }. For scenarios where a macro is indirectly connected to standard cell clusters through other clusters (multi-hop connections), we first compute each cluster’s geometric center and then determine virtual centers based on these for the multi-hop connection. The $V_{mcc}$ vector projection is carried out in the same way, with the results superimposed.

\begin{equation}
\begin{gathered}
x_{geo\_center} = \frac{1}{c} *\sum_{i=1}^{c} x_i ; \ 
y_{geo\_center} = \frac{1}{c} *\sum_{i=1}^{c} y_i
\end{gathered}
    \label{eq: geo center}
\end{equation}

\begin{equation}
    \begin{gathered}
        x \{V_{mc}\} = x\{V_{mc_1}\} + x\{V_{mc_2}\} + ... + x\{V_{mc_n}\} \\
        y \{V_{mc}\} = y\{V_{mc_1}\} + y\{V_{mc_2}\} + ... + y\{V_{mc_n}\}
    \end{gathered}
\label{eq : mc x + y }
\end{equation}

\setlength{\textfloatsep}{0pt}
\begin{algorithm}[t]
    \footnotesize
    \caption{Dataflow decomposition-based macro flipping algorithm (using flipping on x-axis as an example, y-axis is the same)}
    \label{alg:1}
    \renewcommand{\algorithmicrequire}{\textbf{Input:}}
    \renewcommand{\algorithmicensure}{\textbf{Output:}}
    \begin{algorithmic}[1]
        \Require $macros$ $\leftarrow$ pre-placed macros                      
        \Ensure $macros$ $\leftarrow$ macros flipped by dataflow decomposition  

        \For{each macro as center macro}  
            \State $center\_macro\_x = pin\_center\_x$ of center macro
            \State $total\_force\_x = 0$ 
            \State \textcolor{blue}{\Comment{Macro to Macro out degree (Fig. \ref{fig:flip_mm})}}
            \For{each connected macro}  
                \State $macro\_x = pin\_center\_x$ of connected macro
                \State $sub\_force = w1 * (macro\_x - center\_macro\_x)$
                \State $total\_force\_x += sub\_force$
            \EndFor
            \State \textcolor{blue}{\Comment{Macro to Cell Cluster out degree (Fig. \ref{fig:flip_mc} left)}}
            \For{each connected cell cluster}  
                \State $cell\_x = cluster\ geometric\ center$
                \State $sub\_force = w2 * (cell\_x - center\_macro\_x)$
                \State $total\_force\_x += sub\_force$
            \EndFor
            \State \textcolor{blue}{\Comment{Macro to Multi-hop Cell Cluster out degree (Fig. \ref{fig:flip_mc} right)}}
            \For{each connected two-hop cell cluster}  
                \State $cell\_1\_x = cluster\_1\ geometric\ center$
                \State $cell\_2\_x = cluster\_2\ geometric\ center$
                \State $cell\_center\_x = 0.5 * (cell\_1\_x + cell\_2\_x)$
                \State $sub\_force = w3 * (cell\_center\_x - center\_macro\_x)$
                \State $total\_force\_x += sub\_force$
            \EndFor            
        \EndFor \\
        \Return Selected Macros for Flipping
    \end{algorithmic}
\end{algorithm}

\subsubsection{\textbf{Making Flipping Decisions}}
Using the dataflow decomposition methods, we obtain three distinct dataflow vectors: $V_{mm}$, $V_{mc}$, and $V_{mcc}$. The Equation \ref{eq: Vt} represents the weighted sums of these vectors' projections along the x axis and y axis, forming the projections of $V_T$ on these axes:
\begin{equation}
    \begin{gathered}
        x \{V_{T}\} = \alpha * x\{V_{mm}\} + \beta * x\{V_{mc}\} + \gamma * x\{V_{mcc}\} \\
        y \{V_{T}\} = \alpha * y\{V_{mm}\} + \beta * y\{V_{mc}\} + \gamma * y\{V_{mcc}\}
    \end{gathered}
    \label{eq: Vt}
\end{equation}
The weights $\alpha$, $\beta$, and $\gamma$ are hyperparameters that can be adjusted based on the strength of connectivity. In our evaluations, we determined their values based on multiple trials across various design scales, and set $\alpha = 0.55$, $\beta = 0.3$ and $\gamma = 0.15$. With this hyperparameters setting, design delivers a good optimization performance.
In designs with only macro-cell connections, $V_{mc}$ is particularly critical. While $V_{mcc}$ includes $V_{mc}$ cases, direct connections between cell clusters have a more significant impact on macro layout than indirect 2-hop connections, as observed in real physical design \cite{10546560}. Therefore, both $V_{mc}$ and $V_{mcc}$ are considered in our calculations.

\textcolor{black}{The flipping decisions for the macro are made by comparing the magnitudes of $x\{V_T\}$ which guide the left and right flip and $y\{V_T\}$ which guide the up and down flip.  The direction with the larger value indicates a stronger influence of the dataflow in that direction, similar to principal component analysis in linear algebra. The macro is then flipped accordingly.}

\section{Evaluation Results}
\label{evaluation-results}

\subsection{Experiment Setup}
\noindent \textbf{Evaluation Flow.} DAS-MP is implemented in C++ and is compatible with the database used in the latest OpenROAD release \cite{kahng2021openroad,openroad}. To evaluate its capabilities, we tested DAS-MP through the full design flow, including placement and routing. The experiments were conducted on an Intel Core i7-11700 CPU with 128GB of memory. The netlist was generated using the Yosys synthesis tool \cite{wolf2016yosys}. We compare the DAS-MP whole-process approach against Triton Macro Placer (TMP) \cite{tmp}, the default macro placer in OpenROAD, RTL-MP \cite{Kahng2022}, a recently released dataflow-aware macro placer, and Hier-RTLMP \cite{kahng2023hierrtlmp}, a state-of-the-art macro placer in OpenROAD. To better understand the impact of each optimization stage, we evaluate two versions of DAS-MP:
\begin{itemize}
    \item \textbf{DAS-MP (DE)}: Incorporates only the dataflow connection extraction methods \cite{10546560} between macros and standard cells.
    \item \textbf{DAS-MP (DE+FT)}: Includes both dataflow connection extraction and fine-tuning optimizations, such as macro-specificity and macro orientation adjustments.
\end{itemize}


\noindent \textbf{Comparison Metrics.} To evaluate macro placement and fine-tuning quality, we primarily use HPWL, a widely accepted metric in placement and floorplanning research, referenced in recent macro placers \cite{kahng2023hierrtlmp,Kahng2022,Mirhoseini2021,liu2022floorplanning}. Since HPWL alone does not fully reflect the final QoR \cite{kahng2024ppa}, we also assess post-routing PPA metrics, including WNS, TNS, power, and area. Runtime is recorded to evaluate the framework’s efficiency.\par

\noindent \textbf{Benchmark Selection.} We select the benchmarks based on the following criteria. Firstly, they need to be representative and widely accepted in the community. Secondly, they should exhibit diverse behaviors related to macros. For example, some designs should be macro-dominant, and some should have more connections among macros and cell clusters. Last but not least, they need to be new enough to reflect the current design scale and complexity. Based on the above principles, we run extensive experiments and end up picking seven benchmark designs for the evaluation of the proposed methodology. They are selected based on commonly used design cases in recently published literature or newly released test suites for evaluating macro placement \cite{tilos,openroad}. We test the designs in  NanGate45 \cite{Nangate} technology node, a popularly used open-source standard cell libraries in 45nm.


\begin{table*}[ht]
\centering
\caption{Benchmark Information And Extracted Detailed Connections for Each Benchmark Design}
\label{Tab:design information}
\resizebox{\textwidth}{!}{
\begin{threeparttable}
    \begin{tabular}{l|ccccc|ccccc}
        \toprule
        Design\ Name &  \makecell{Std Cells \\Count} &  \makecell{Macros \\Count} & \makecell{Macro\\Type \tnote{1} } & \makecell{\# of \\ IOs}& \makecell{Total Cluster\\ Count} & \textbf{\makecell{Macro Cluster- \\ Macro Cluster \tnote{*}}} & \textbf{\makecell{Macro Cluster- \\ Cell Cluster \tnote{*} \  \tnote{2} }}  & \textbf{\makecell{Cell Cluster- \\ Macro Cluster \tnote{*} \ \tnote{2} }}  &\textbf{\makecell{Cell Cluster- \\ Cell Cluster \tnote{*}}}  & \textbf{\makecell{Macro Cluster- \\ Cell Cluster-Cell Cluster \tnote{*}}} \\ 
        \midrule
        \texttt{TinyRocket} & 27217 & 2 & 1 & 269& 65 &0 & 2 & 35 & 829 & 67\\ 
        \texttt{bp\_be} & 59882 & 10 & 3 &  3029& 139 &25 & 120&364 &1815 & 251 \\ 
        \texttt{bp\_fe} & 29993 & 11 & 3 &  2511& 82 &24 &132 &260 & 406& 196\\
        \texttt{black parrot} & 427501 & 24 & 5 &  1198 & 763 & 2 & 565 & 1095 & 14649&3375\\ 
        \texttt{bp\_multi}  & 209086 & 26 & 6 &  1453& 432 &2 & 351 & 828& 4962 &1427 \\ 
        \texttt{swerv\_wrapper} & 99750 & 28 & 3 &  1416& 518 &0&232&741&15300 & 594 \\ 
        \texttt{ariane133} & 165953 & 133 & 1  & 495& 314 & 44 &380 &1947 &9963 &1708  \\ 
        
 \bottomrule
    \end{tabular}
\begin{tablenotes}
        \item [*] \# of unique connections. 
        \item [1] This includes macros with different sizes and functionalities.
        \item [2] Bi-directional connections.
        \end{tablenotes}
  \end{threeparttable}}
  \vspace{-10pt}
\end{table*}

\subsection{Dataflow Connection Relationship Analysis}
Table \ref{Tab:design information} summarizes the benchmark characteristics and the number of unique dataflow connections extracted using DAS-MP. \texttt{TinyRocket} is a small design with only two macros of the same type and a limited number of standard cells, but it exhibits a high number of {\tt cell cluster-cell cluster} connections. \texttt{bp\_multi} has a macro count comparable to \texttt{black\ parrot}, but their dataflow patterns differ significantly. \texttt{bp\_multi} has substantially fewer {\tt macro cluster-cell cluster} and {\tt cell cluster-macro cluster} connections than \texttt{black\ parrot}, with an even greater disparity in {\tt cell cluster-cell cluster} connections. \texttt{Ariane133}, selected from the macro placement test suite \cite{tilos}, represents a macro-heavy design but features only one unique macro type, resulting in fewer unique connections. For instance, its number of macro cluster-cell cluster connections is lower than that of \texttt{black\ parrot}. This analysis enables more design-specific macro placement, which is particularly valuable when implementing a new design without prior experience.

\begin{figure}[!t]
\vspace{-10pt}
    \centering
    \subfigure[RTL-MP]{
    \label{fig:back_parrot_rtlmp}
    \includegraphics[width=0.3\linewidth]{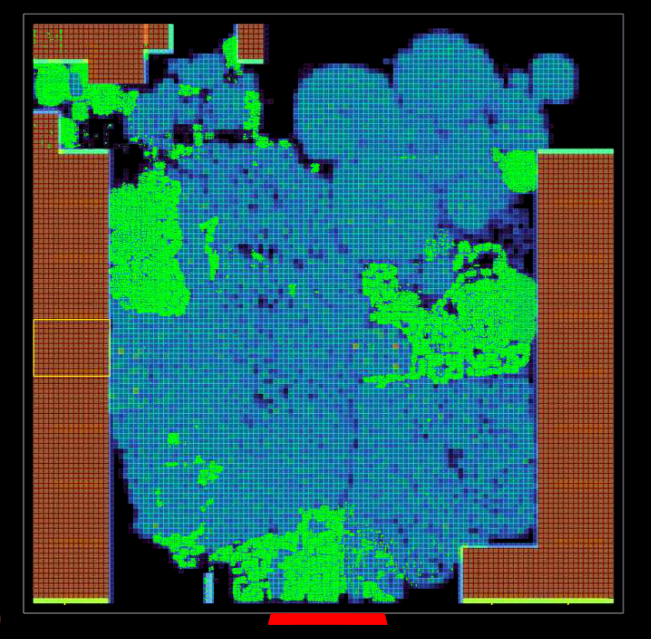}
    }
    \subfigure[DAS-MP (DE)]{
    \label{fig:back_parrot_dasmp}
    \includegraphics[width=0.3\linewidth]{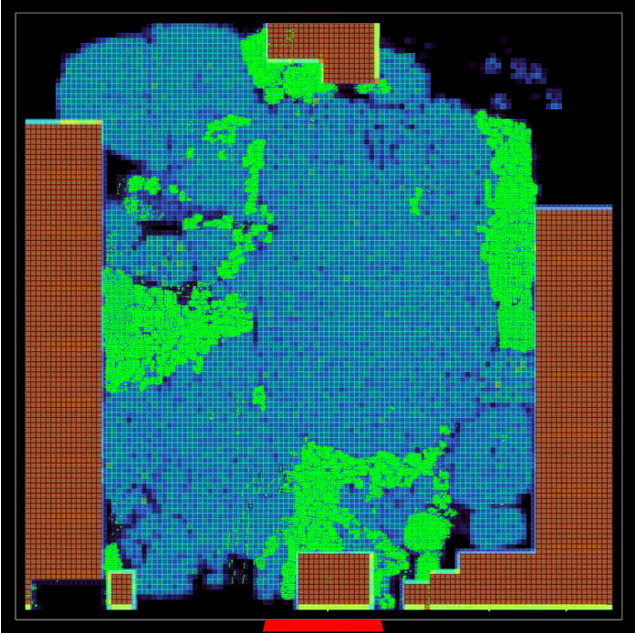}
    }
    \caption{Layouts of {\tt black\ parrot} with congestion map. The highlighted green cells are from the one-hop connected cell cluster. }
    \label{fig:back_parrot}
\end{figure}

\begin{table*}
\vspace{-5pt}
\caption{Experimental Results}
\centering
\label{tab:experimental result}
\resizebox{1.0\textwidth}{!}{
\begin{threeparttable}
\begin{tabular}{l|cc|cc|cc|cc|cc|cc} 
\toprule[1pt]
\multirow{3}{*}{Design Name} & \multicolumn{2}{c}{TMP \cite{kahng2021openroad}}             & \multicolumn{2}{c}{RTL-MP\cite{Kahng2022} (\textbf{baseline})}   & \multicolumn{2}{c}{Hier-RTLMP\cite{kahng2023hierrtlmp}   }        & \multicolumn{4}{c}{DAS-MP (DE) \cite{10546560} (Proposed)} & \multicolumn{2}{c}{\textbf{DAS-MP (DE+FT) (Proposed)}}             \\
\cmidrule[1pt]{2-13}
                           & HPWL (m) &\makecell{Congestion \\ Overflow}  & HPWL (m) &\makecell{Congestion \\ Overflow}  & HPWL (m) &\makecell{Congestion \\ Overflow}  & \makecell{\textbf{M-C } \\ \textbf{HPWL (m)}}   &\makecell{\textbf{Congestion}\\ \textbf{Overflow}}&\makecell{\textbf{M-C-C}\\ \textbf{HPWL (m)}}& \makecell{\textbf{Congestion}\\ \textbf{Overflow}}  & \textbf{HPWL (m)} &\makecell{\textbf{Congestion} \\ \textbf{Overflow}} \\ 
\midrule[1.2pt]
\texttt{TinyRocket}                 & 830.03    &  73   & 803.80 &   74    & 786.75     &   \textbf{0}   &  758.98 &26 &  752.80  &19 &\textbf{744.29}  & 16\\
\texttt{bp\_be }                    & 4151.71 &66553   & 4445.60  & 40537     & 4798.22       &  3738    &  4218.92 & 2224 &4113.33  &454 &\textbf{4003.06}  &\textbf{433}  \\
\texttt{bp\_fe}                     & 2772.04 & 3373      &   2789.94 & 4445    & 3130.68      & \textbf{0}  &  2664.61 & 3165 & 2519.63 & 3156 &\textbf{2463.79}  & 1530\\
\texttt{black parrot}               &12966.52&  142       & 12930.57& 981    & 13430.91   & 1570  &12600.53  & 272  &11820.00 & 130  &\textbf{11632.06}  &\textbf{115}\\
\texttt{bp\_multi}                  & 7496.04& 2854       & 7252.76& 3965     & 7630.34  & 62699 & 7235.15 &  1328 & 7146.48   & 1470 &\textbf{6999.51}  & \textbf{173}\\
\texttt{swerv\_wrapper}              &5199.62& 3384          &   4745.86 &2533  &5400.98    & \textbf{633}   & 4718.80    &2299  & 4648.02 & 887  & \textbf{4366.34}  & 848   \\
\texttt{ariane133}                  & 8775.61& 3421       & 8857.12& 23876  &9525.40  &  183261         & 8737.35 & 3842   &   8624.48  &3919 &\textbf{8378.74}  &\textbf{3707}  \\ 
\midrule[1pt]
\textbf{Avg. Improvement}&  0.9\% \textcolor{blue}{\textcolor{blue}{$\downarrow$}}  & 4.4\% \textcolor{blue}{$\downarrow$}       &    0\% & 0\%   & 6.9\% \textcolor{blue}{$\downarrow$} & 26.3\% \textcolor{blue}{$\downarrow$} & 2.8\%\textcolor{red}{\textcolor{red}{$\uparrow$}} & 62.9\% \textcolor{red}{$\uparrow$}  & 5.5\% \textcolor{red}{$\uparrow$}  & 73.4\% \textcolor{red}{$\uparrow$}  & \textbf{7.9\%} \textcolor{red}{$\uparrow$} & \textbf{82.5\%} \textcolor{red}{$\uparrow$} \\ 
\bottomrule[1.2pt]
\end{tabular} 
  \end{threeparttable}}
\end{table*}

\subsection{HPWL Result Analysis}

Table \ref{tab:experimental result} presents HPWL results for TMP \cite{kahng2021openroad}, RTL-MP \cite{Kahng2022}, Hier-RTLMP \cite{kahng2023hierrtlmp}, DAS-MP (DE) with dataflow extraction only, and DAS-MP (DE+FT) with full optimization. The best result for each design is highlighted in bold, with percentage improvements over RTL-MP provided, as RTL-MP generally outperforms TMP and Hier-RTLMP.

\subsubsection{HPWL Improvement from Dataflow-aware Placement (w/o Fine-tuning)}
The proposed DAS-MP (DE) method outperforms RTL-MP across all designs, achieving HPWL improvements ranging from 0.2\% to 9.7\%. Using \texttt{black\ parrot} as an example, as illustrated in Fig. \ref{fig:back_parrot_rtlmp} and Fig. \ref{fig:back_parrot_dasmp}, DAS-MP (DE) achieves lower HPWL by placing cells and macros closer together. Notably, the previous macro placer \cite{Kahng2022} failed to recognize certain critical cell cluster connections near the pin access region, leading to increased wire length. Our method corrects this by accurately identifying these connections and adjusting macro placement accordingly. While blocking pin access may seem counterintuitive, our experiments show that it can sometimes enhance placement quality. In simpler designs with fewer macros, macro placement is generally more straightforward, but overlooking macro-cell connections can still result in unnecessary design iterations. Additionally, automatic macro placers can be misled by poorly defined constraints, a challenge DAS-MP (DE) overcomes with its dataflow-guided approach.


\begin{figure}[t]
    \centering
    \subfigure{
    \includegraphics[width=1.0 \linewidth]{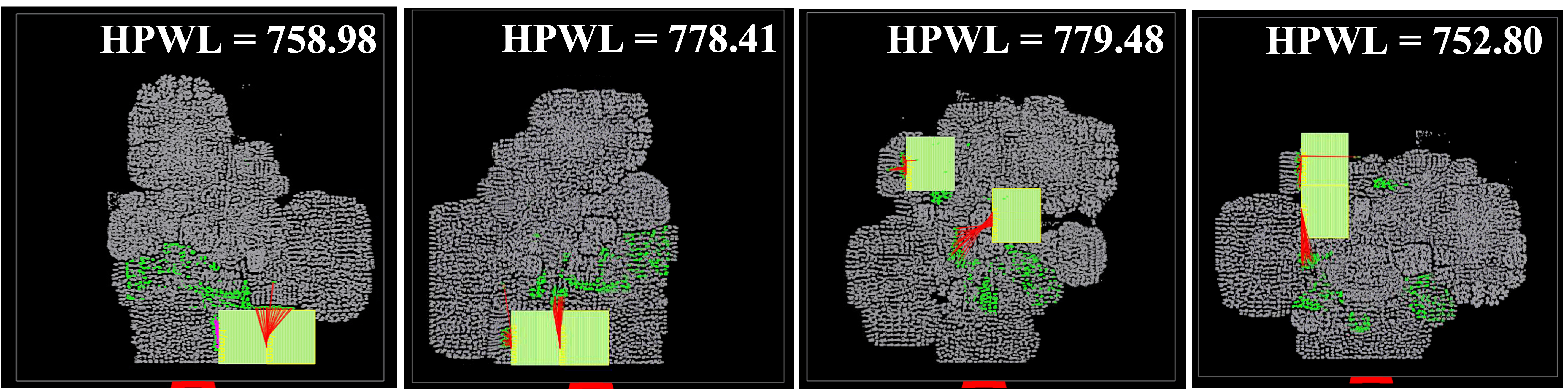}
    }
    \vspace{15pt}
\resizebox{0.9\columnwidth}{!}{
\begin{tabular}{lcc}
\toprule[1pt]
                                  & \makecell{Macro Cluster-Cell Cluster \\ HPWL (m)} & \makecell{Macro Cluster-Cell Cluster-\\Cell Cluster HPWL (m)}\\ \midrule[1pt] 
\multicolumn{1}{l|}{Push boundary}     & 758.98       & 778.41            \\
\multicolumn{1}{l|}{Not push boundary} & 779.48       & \textbf{752.80}   \\ 
\bottomrule[1pt]
\end{tabular}
}
\vspace{-10pt}
    \caption{The proposed method placement result of {\tt TinyRocket} with ``push boundary'' and ``not push boundary'' actions.}
\label{fig:tinyrocket boundary}
\end{figure}

\subsubsection{HPWL Improvement from Dataflow-aware Placement (w/ Fine-tuning)}
The right two columns of Table \ref{tab:experimental result} present the HPWL and congestion overflow results of DAS-MP (DE+FT), which incorporates fine-tuning on top of the dataflow extraction-only strategy. The results demonstrate that fine-tuning further improves HPWL, achieving an average reduction of 7.9\% compared to RTL-MP across all designs, while also significantly reducing congestion overflow by 82.5\% on average. Notably, all designs outperform the dataflow-only approach (DAS-MP (DE)). The improvement is particularly evident in designs such as \texttt{bp\_fe} and \texttt{swerv\_wrapper}, which achieve HPWL reductions of 2.8\% and 6.1\%, respectively. In \texttt{bp\_be}, fine-tuning reduces HPWL by approximately 9\% compared to RTL-MP, optimizing wirelength by bringing cells and macros closer together while minimizing unnecessary routing. Similarly, \texttt{swerv\_wrapper} shows significant gains in both HPWL and congestion overflow, confirming that fine-tuning substantially enhances placement quality, especially in designs with more complex connectivity.
\begin{figure*}[t]
    \centering
    \subfigure[TMP \cite{kahng2021openroad}]{
    \label{bpfetmp}
    \includegraphics[width=0.18\linewidth]{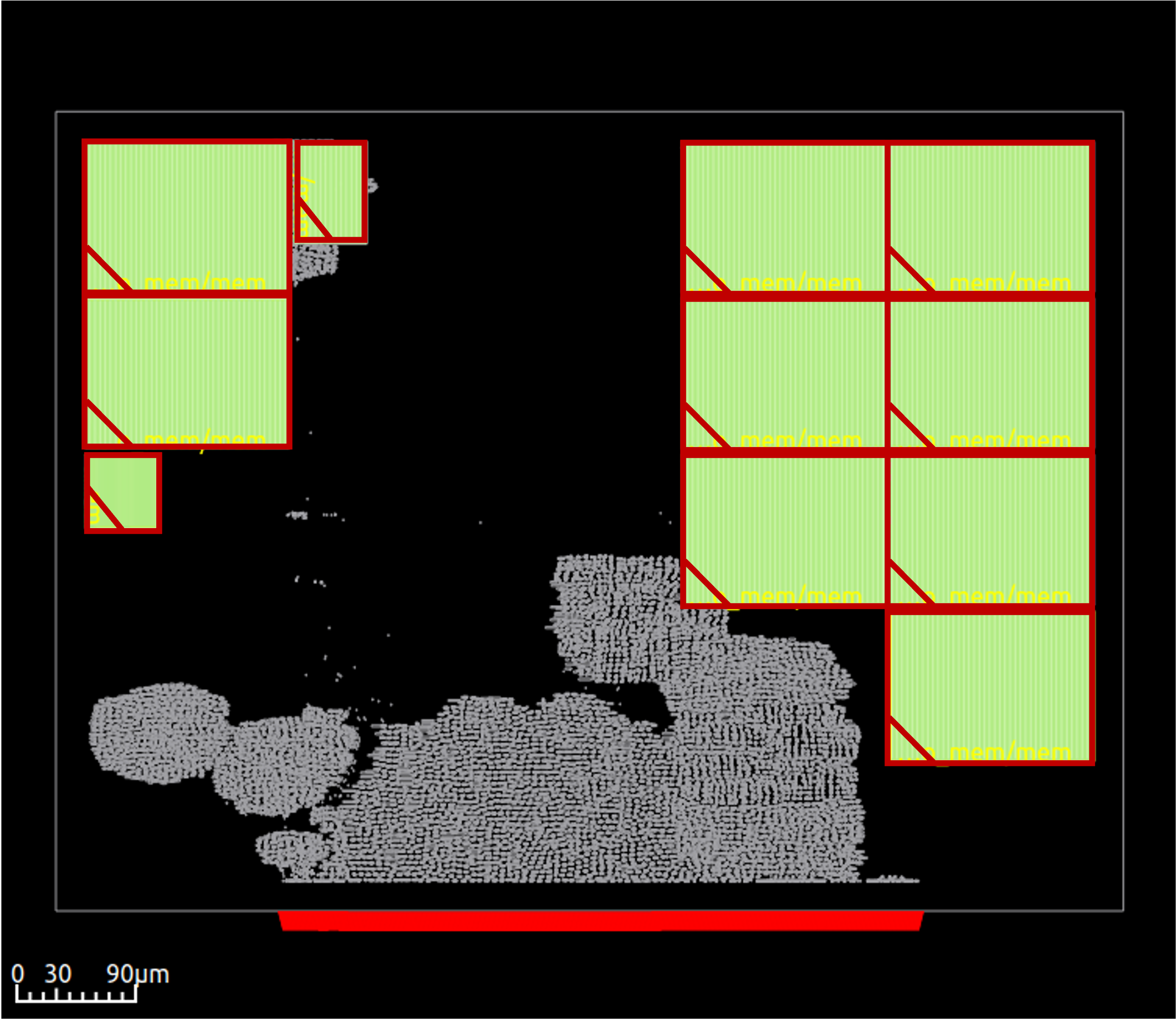}
    }
    \subfigure[RTL-MP \cite{Kahng2022}]{
    \label{bpfertlmp}
    \includegraphics[width=0.18\linewidth]{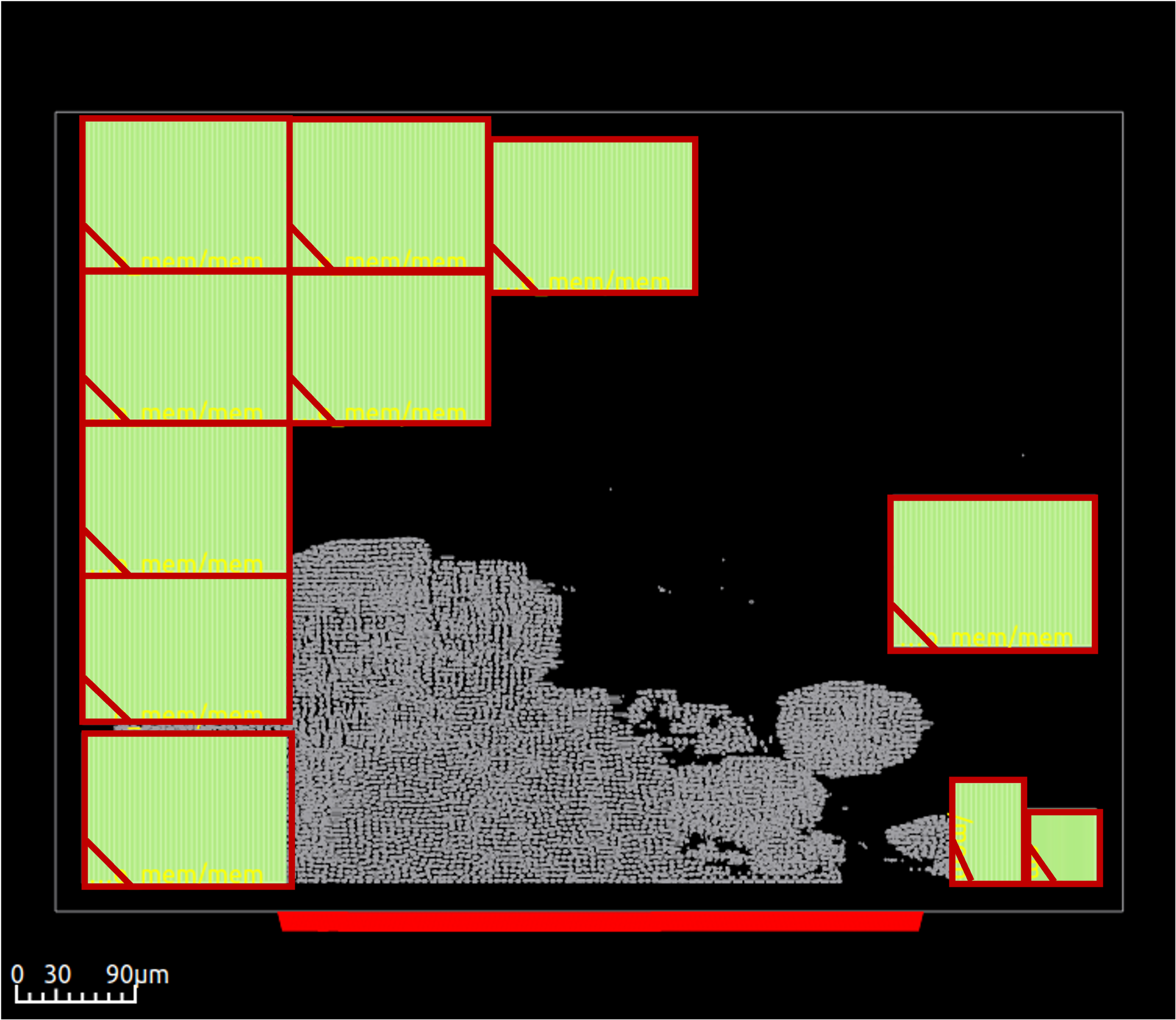}
    }
    \subfigure[Hier-RTLMP \cite{kahng2023hierrtlmp}]{
    \label{bpfehier}
    \includegraphics[width=0.18\linewidth]{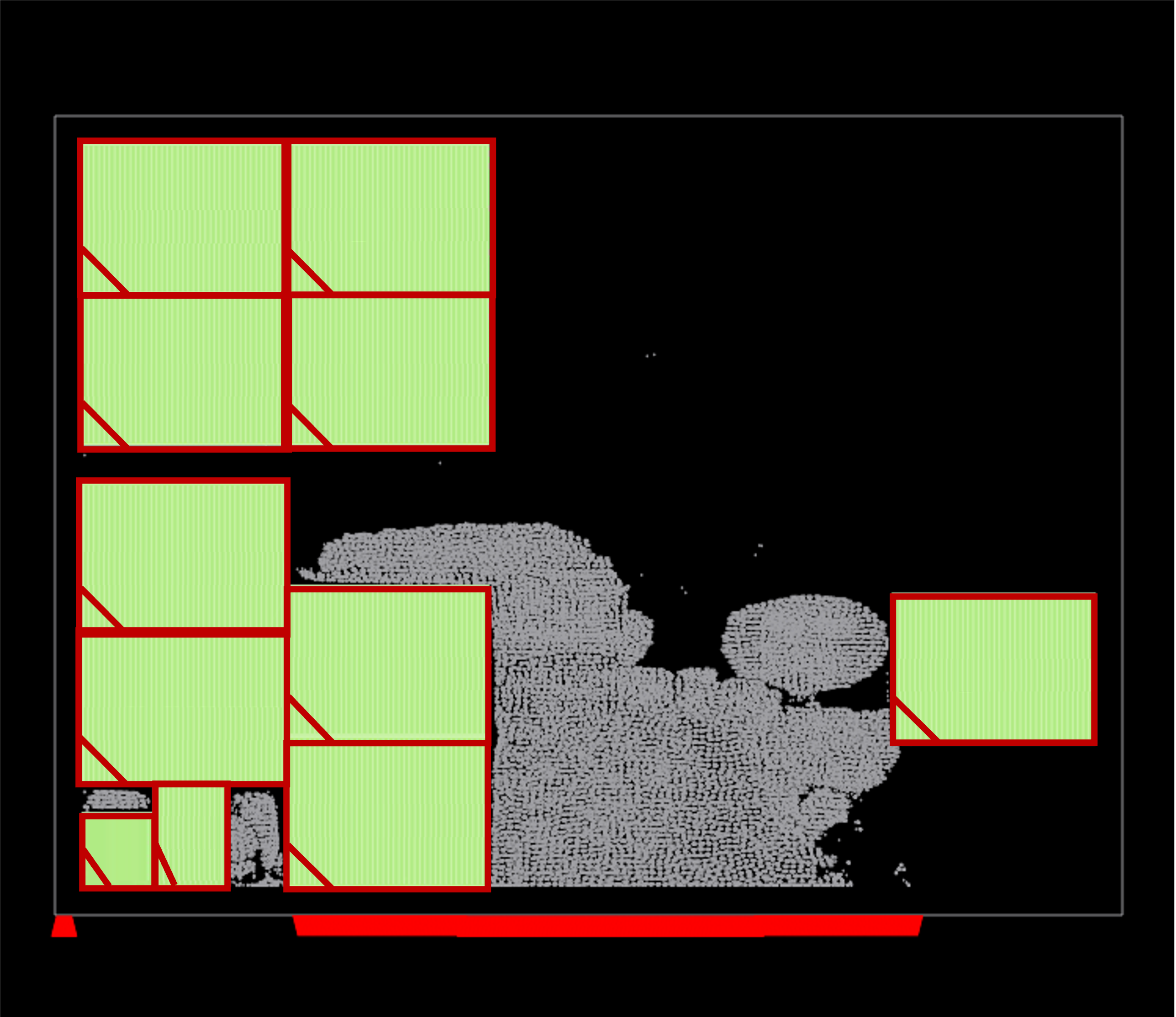}
    }
    \subfigure[DAS-MP (DE) \cite{10546560}]{
    \label{bpfede}
    \includegraphics[width=0.18\linewidth]{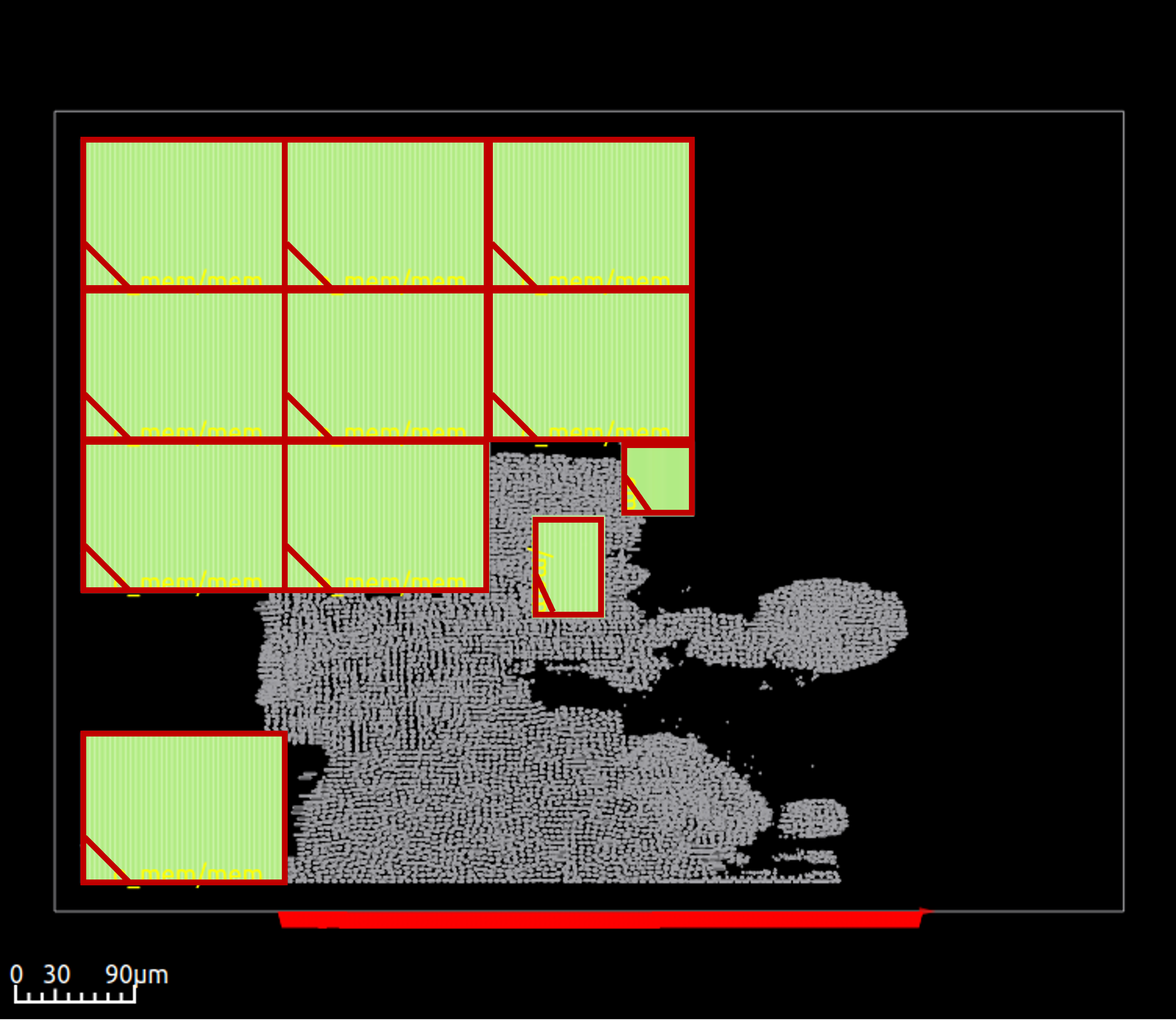}
    }
    \subfigure[DAS-MP (DE+FT)]{
    \label{bpfeft}
    \includegraphics[width=0.18\linewidth]{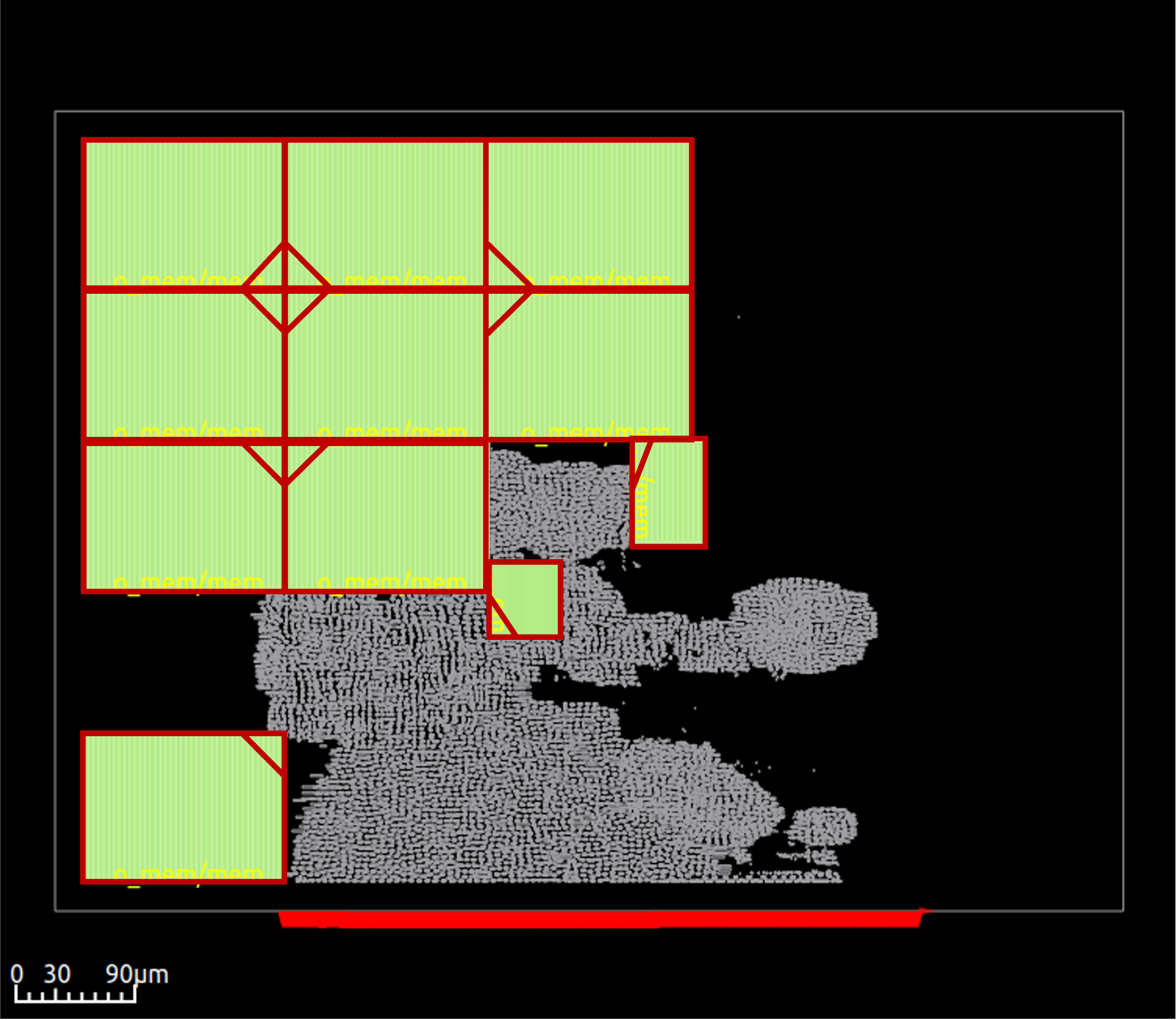}
    }
    \caption{Layouts of different macro placers for {\tt bp\_fe}. The red triangles on the corner indicate the orientation of each macro. }
    \label{fig:bpfe layout}
\end{figure*}

\begin{figure*}[t]
\vspace{-5pt}
    \centering
    \subfigure[TMP \cite{kahng2021openroad}]{
    \label{swtmp}
    \includegraphics[width=0.18\linewidth]{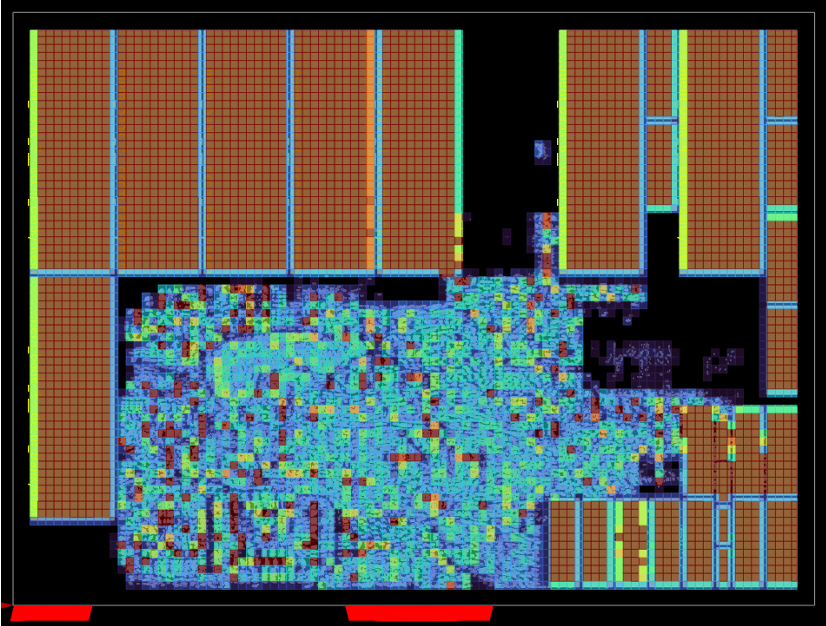}
    }
    \subfigure[RTL-MP \cite{Kahng2022}]{
    \label{swrtlmp}
    \includegraphics[width=0.185\linewidth]{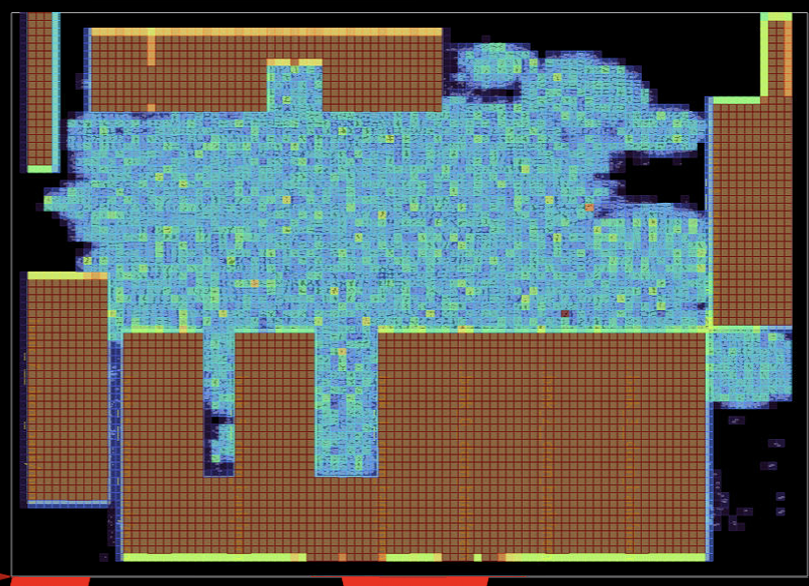}
    }
    \subfigure[Hier-RTLMP \cite{kahng2023hierrtlmp}]{
    \label{swhier}
    \includegraphics[width=0.18\linewidth]{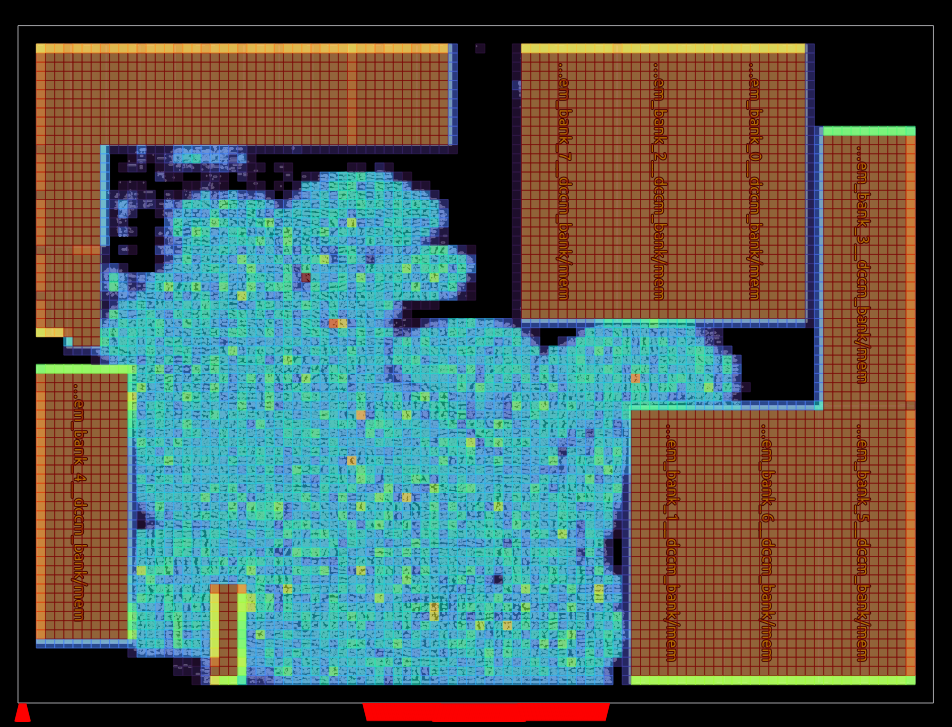}
    }
    \subfigure[DAS-MP (DE) \cite{10546560}]{
    \label{swde}
    \includegraphics[width=0.18\linewidth]{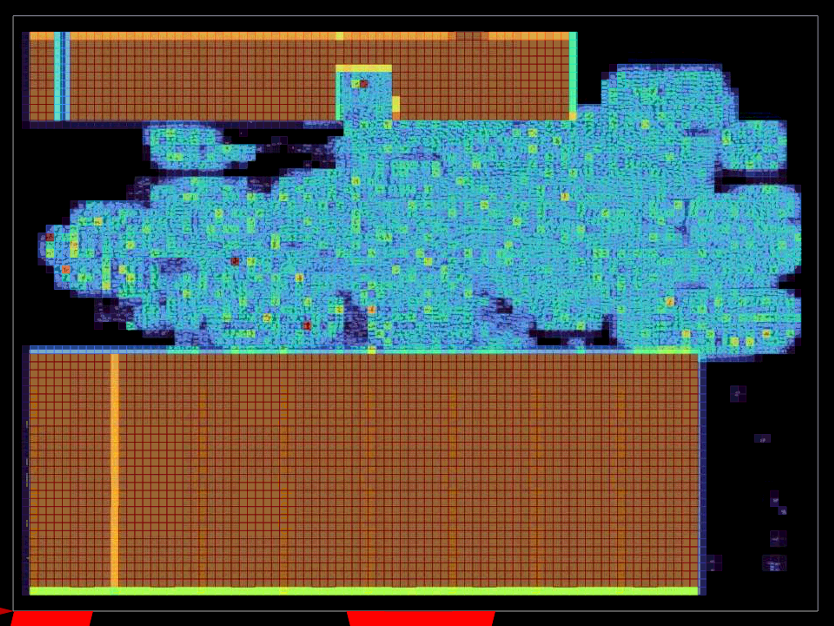}
    }
    \subfigure[DAS-MP (DE+FT)]{
    \label{swft}
    \includegraphics[width=0.176\linewidth]{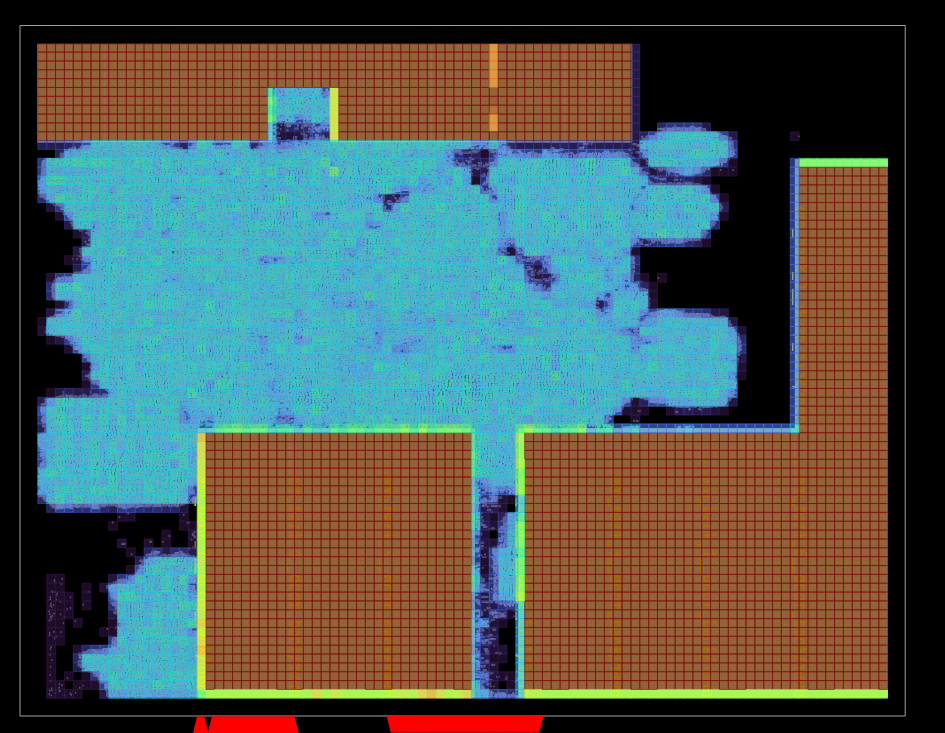}
    }
    \caption{Congestion maps of different macro placers for {\tt swerv\_wrapper}.}
    \label{fig:sw layout}
    \vspace{-5pt}
\end{figure*}

\subsubsection{The Impact of ``Push Boundary'' Action}
\label{The Impact of Pushing Boundary Action}
A common practice among physical designers is to push macros closer to the boundary to create more space for standard cells. This approach assumes that standard cells are typically placed after macro locations are fully or partially fixed. However, DAS-MP reveals that this ``rule'' is not always optimal. Stronger dataflow connections can increase wirelength if macros are pushed to the edge, far from the connected standard cells. Therefore, DAS-MP relaxes this constraint and observes the resulting placement quality. An example is shown in Fig. \ref{fig:tinyrocket boundary} using $\texttt{TinyRocket}$, the simplest design in the suite, which is cell-dominant and has a large number of {\tt cell cluster-cell cluster} connections. The table summarizes the HPWL differences when enabling or disabling the pushing-boundary constraint. When considering only one-hop connections, ``pushing boundary'' results in better HPWL due to fewer cell clusters involved. However, for two-hop connections, ``not pushing boundary'' achieves better HPWL, leading to an overall 6.3\% improvement over RTL-MP. This demonstrates that DAS-MP can uncover optimization opportunities that may be overlooked by conventional design practices.


\subsection{Congestion Overflow Result Analysis}

By considering ``hidden'' connections, congestion conditions can be significantly improved. This improvement is reflected in the reduced congestion overflow reported in Table \ref{tab:experimental result}. The proposed DAS-MP (DE+FT) achieves an average congestion overflow reduction of 82.5\%, outperforming all other compared placers. This is due to the placement of macros and their associated cell clusters in closer proximity, which helps avoid long and zigzag wires. Additionally, macros are adjusted and flipped to their optimal orientations, further reducing wirelength and improving routing efficiency. As a result, more routing resources become available, leading to overall better placement quality.

Since \texttt{bp\_fe} and \texttt{swerv\_wrapper} demonstrated relatively significant optimization, we take them as examples. Table \ref{tab:experimental result} shows that DAS-MP (DE+FT) not only achieves superior HPWL results but also excels in other quality metrics. This is further illustrated in Fig. \ref{fig:bpfe layout} and Fig. \ref{fig:sw layout}, where the placement results and congestion maps are presented. It is evident that DAS-MP (DE+FT) effectively places cells closer to their associated macros, leveraging extracted connections to guide the placement of both types of instances. The macro placement achieved by DAS-MP (DE+FT) closely aligns with manual efforts, where macros belonging to the same hierarchy are typically placed in clusters and oriented towards the dataflow, leading to improved legalization. The congestion map in Fig. \ref{fig:sw layout} further highlights the advantage of DAS-MP, as it produces a layout with fewer congestion hotspots compared to the other methods. Additionally, DAS-MP improves resource utilization, leaving more empty space available for subsequent optimizations in detailed placement, routing, and other stages of the design flow.


\subsection{Post-routing PPA Result Analysis}

\begin{table}[!t]
\centering
\caption{Post-Routing PPA Results}
\label{tab: PPA experimental result}
\resizebox{\columnwidth}{!}{
\begin{threeparttable}
\begin{tabular}{c|l|c|c|c|c} 
\toprule[1.2pt]
\textbf{Design}         & \textbf{Flow}       & \textbf{WNS} (ns)    & \textbf{TNS} (ns)       & \textbf{Power} (mw) & \textbf{Area}  ($\mu m^2$)        \\\midrule[1.2pt]
\multirow{5}{*}{\texttt{TinyRocket}}     
                & TMP\cite{kahng2021openroad}       & -0.270  & -88.695   &  74    &   59528   \\
               & RTL-MP\cite{Kahng2022}     & -0.259 & -83.636   &   \textbf{73}   &  59538  \\
               & Hier-RTLMP\cite{kahng2023hierrtlmp} & -0.230 & -69.035    &   191    &    \textbf{59510}       \\
               & DAS-MP (DE) (M-C)\cite{10546560}       & -0.236 & -73.763   &   74     &    59536     \\
               & DAS-MP (DE) (M-C-C)\cite{10546560}      & -0.228 & -68.877   &   \textbf{73}     &  59539       \\ 
               & \textbf{DAS-MP (DE+FT)} &\textbf{-0.220}   & \textbf{-68.533}   &   \textbf{73}     &  59535     \\ 
               \midrule[1pt]
\multirow{5}{*}{\texttt{bp\_be} }       
               & TMP\cite{kahng2021openroad}        &  -1.081 &  -1238.863   &    419  &   269138    \\
               & RTL-MP\cite{Kahng2022}    & -0.830 & -1005.879  &     414      &    269204       \\
               & Hier-RTLMP\cite{kahng2023hierrtlmp} & -1.139 & -1237.849   &   410     &     \textbf{268252}    \\
               & DAS-MP (DE) (M-C)\cite{10546560}      & -0.638 &  -667.922 &  423  &      269116     \\
               & DAS-MP (DE) (M-C-C)\cite{10546560}      & -0.618 & -467.990 &   411    &     268987    \\
               & \textbf{DAS-MP (DE+FT)}     & \textbf{-0.610}   & \textbf{-313.790}    &  \textbf{389}     & 268357        \\
               \midrule[1pt]
\multirow{5}{*}{\texttt{bp\_fe}  }       
               & TMP\cite{kahng2021openroad}        & -0.533 & -27.161   &     275   &      222466     \\
               & RTL-MP\cite{Kahng2022}    & -0.702 & -37.492   &     264  &     221949     \\
               & Hier-RTLMP\cite{kahng2023hierrtlmp} & -3.388 & -9802.706 &     1010   &   212604      \\
               & DAS-MP (DE) (M-C)\cite{10546560}      & -0.717 & -42.103   &     \textbf{259}   &    \textbf{221769}     \\
               & DAS-MP (DE) (M-C-C)\cite{10546560}      & -0.531 & -22.327   &     \textbf{259}   &    221774      \\
               & \textbf{DAS-MP (DE+FT)}       &\textbf{-0.487}   & \textbf{-17.855}    & 270       & 222251      \\
               \midrule[1pt]
\multirow{5}{*}{\texttt{black parrot} }  
               & TMP\cite{kahng2021openroad}        & -0.108 & -0.108    & 168       & 877591   \\
               & RTL-MP\cite{Kahng2022}    & -0.147 & -0.147    & 169       & 877590   \\
               & Hier-RTLMP\cite{kahng2023hierrtlmp} & -0.119 & -0.119    & 206       & \textbf{877248}   \\
               & DAS-MP (DE) (M-C)\cite{10546560}      & -0.128 & -0.128    & 157       & 877563   \\
               & DAS-MP (DE) (M-C-C)\cite{10546560}      & \textbf{-0.095} & \textbf{-0.095}    & \textbf{154}       & 877477  \\
               & \textbf{DAS-MP (DE+FT)}       & -0.110  & -0.110    &  169      &  877803     \\
               \midrule[1pt]
\multirow{5}{*}{\texttt{bp\_multi}  }    
                & TMP\cite{kahng2021openroad}        & -0.223 & -77.322   & 218       & 614172   \\
               & RTL-MP\cite{Kahng2022}    & -0.25  & -121.213  & 218       & 614173        \\
               & Hier-RTLMP\cite{kahng2023hierrtlmp} & -0.261 & -105.773  & 254       & \textbf{613882}   \\
               & DAS-MP (DE) (M-C)\cite{10546560}      & -0.241 & -83.441   & 214       & 615344        \\
               & DAS-MP (DE) (M-C-C)\cite{10546560}      & -0.214 & -75.588   & \textbf{207}       & 615342        \\
               & \textbf{DAS-MP (DE+FT)}       & \textbf{-0.21}  &  \textbf{-0.21}    & 218       & 614539      \\
               \midrule[1pt]
\multirow{5}{*}{\texttt{swerv\_wrapper}} 
               & TMP\cite{kahng2021openroad}        & -1.098 & -1441.25  &  26600    &     511702    \\
               & RTL-MP\cite{Kahng2022}    & -1.831 & -3063.155 &  26600    &     \textbf{511649}    \\
               & Hier-RTLMP\cite{kahng2023hierrtlmp} & -1.563 &  -2563.767  &  26600    &   511694       \\
               & DAS-MP (DE) (M-C)\cite{10546560}      & -0.958 & -2035.965 &  27600 &     511798      \\
               & DAS-MP (DE) (M-C-C)\cite{10546560}      & -0.853 & -1793.158 &   26600  &   511691      \\
               & \textbf{DAS-MP (DE+FT)}       & \textbf{-0.798}  & \textbf{-1403.92}    &  \textbf{26600}      &  511673     \\
               \midrule[1pt]
\multirow{5}{*}{\texttt{Ariana133}    }  
                & TMP\cite{kahng2021openroad}        & -0.199 & -16.54    &  207    &  739424       \\
               & RTL-MP\cite{Kahng2022}    & -0.109 & -2.26     & 187       & \textbf{739303}        \\
               & Hier-RTLMP\cite{kahng2023hierrtlmp} & -0.056 & -0.859    & 422       & 738914        \\
               & DAS-MP (DE) (M-C)\cite{10546560}      &  -0.076  &  -1.364  &    186   & 739451        \\
               & DAS-MP (DE) (M-C-C)\cite{10546560}      & -0.014 &   -0.063  &     \textbf{186}   & 739408        \\
               & \textbf{DAS-MP (DE+FT)}       & \textbf{-0.012}  & \textbf{-0.053 }   &224&  739307     \\
\midrule[1pt]
\multirow{5}{*}{\textbf{Avg. Improv.}\tnote{*}}
               & TMP\cite{kahng2021openroad}        & 2.23\%\textcolor{blue}{$\downarrow$} & 73.97\%\textcolor{blue}{$\downarrow$}   &  2.37\% \textcolor{blue}{$\downarrow$}   &  0.03\% \textcolor{blue}{$\downarrow$}      \\
               & RTL-MP\cite{Kahng2022}   &  0.00\%  & 0.00\%   & 0.00\%      &  0.00\%        \\
               & Hier-RTLMP\cite{kahng2023hierrtlmp} & 47.25\%\textcolor{blue}{$\downarrow$} & 3705.95\%\textcolor{blue}{$\downarrow$}    & 85.56\%\textcolor{blue}{$\downarrow$}       & \textbf{0.68\%} \textcolor{red}{$\uparrow$}         \\
               & DAS-MP (DE) (M-C)\cite{10546560}      &  17.77\% \textcolor{red}{$\uparrow$}  &  21.48\%\textcolor{red}{$\uparrow$}  &    0.66\% \textcolor{red}{$\uparrow$}  & 0.017\% \textcolor{blue}{$\downarrow$}      \\
               & DAS-MP (DE) (M-C-C)\cite{10546560}      & 36.03\% \textcolor{red}{$\uparrow$}&   46.18\% \textcolor{red}{$\uparrow$} &     \textbf{2.48\%} \textcolor{red}{$\uparrow$} & 0.006\% \textcolor{blue}{$\downarrow$}       \\
               & \textbf{DAS-MP (DE+FT)}      & \textbf{36.97\%} \textcolor{red}{$\uparrow$} & \textbf{59.44\%} \textcolor{red}{$\uparrow$}    & 2.24\% \textcolor{blue}{$\downarrow$}  &   1.3\%\textcolor{blue}{$\downarrow$}     \\
\bottomrule[1.2pt]
 \end{tabular}
 \begin{tablenotes}
        \item [*] The TNS and WNS numbers are always negative, and the closer they are to 0, the more effective the placement.
        In this table, improvements are derived using actual data, which includes negative numbers. 
        Therefore, negative improvement values of TNS and WNS indicate better performance of the method.
    \end{tablenotes}
\end{threeparttable}}
\end{table}

The optimization of HPWL can have a direct impact on timing. In most cases, reducing wirelength leads to shorter signal propagation delays, which in turn helps to improve timing metrics such as Total Negative Slack (TNS) and Worst Negative Slack (WNS). Table \ref{tab: PPA experimental result} summarizes the PPA results after routing with various placement strategies. Since TNS and WNS are conventionally expressed as negative values, calculations are based on their actual (negative) values. A positive ``improvement'' value indicates an increase in TNS or WNS, signifying a degradation in performance. Conversely, a negative ``improvement'' value denotes a reduction in TNS or WNS, indicating an enhancement in timing performance.

Table \ref{tab: PPA experimental result} clearly demonstrates a significant increase in both WNS and TNS for TMP and Hier-RTLMP compared to RTL-MP, indicating that their timing issues have worsened. On the other hand, WNS and TNS significantly decrease (in absolute terms) for DAS-MP (DE) with the 1-hop and 2-hop methods, as well as for DAS-MP (DE+FT). The proposed DAS-MP (DE+FT) method achieves an average WNS improvement of 36.97\%, while TNS improves by an average of 59.44\%. When comparing the HPWL data, it becomes evident that the substantial reduction in HPWL achieved by the proposed methods aligns logically with the significant improvements observed in TNS and WNS, reinforcing the correlation between wirelength and timing optimization.

Since the primary focus of this work is HPWL optimization, extensive changes for area and power results have been avoided. Nonetheless, experimental results indicate that the approach does not negatively impact area and power metrics. The increase in area is minimal, amounting to only 1.3\%, with a corresponding power increase of 2.24\%.




The analysis of the PPA experimental results reveals that the proposed method demonstrates stronger performance in terms of TNS and WNS while avoiding significant negative effects on power and area.

\subsection{Optimization Runtime Analysis}
\begin{figure}[t]
    \centering
    \subfigure[Runtime occupation ratio on the {\tt swerv\_wrapper} (left) and {\tt ariane133} (right) test cases.]{
    \includegraphics[width=0.9\linewidth]{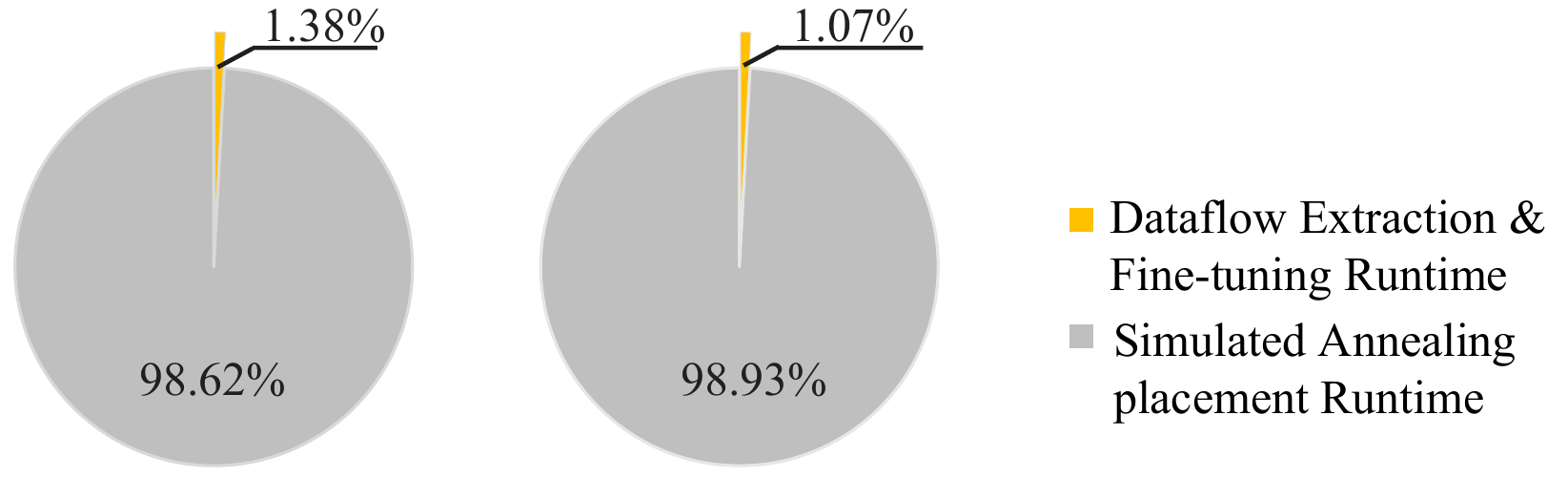}
    \label{fig:runtime1}
    }
    \subfigure[{\tt swerv\_wrapper} breakdown]{
    \includegraphics[width=0.45\linewidth]{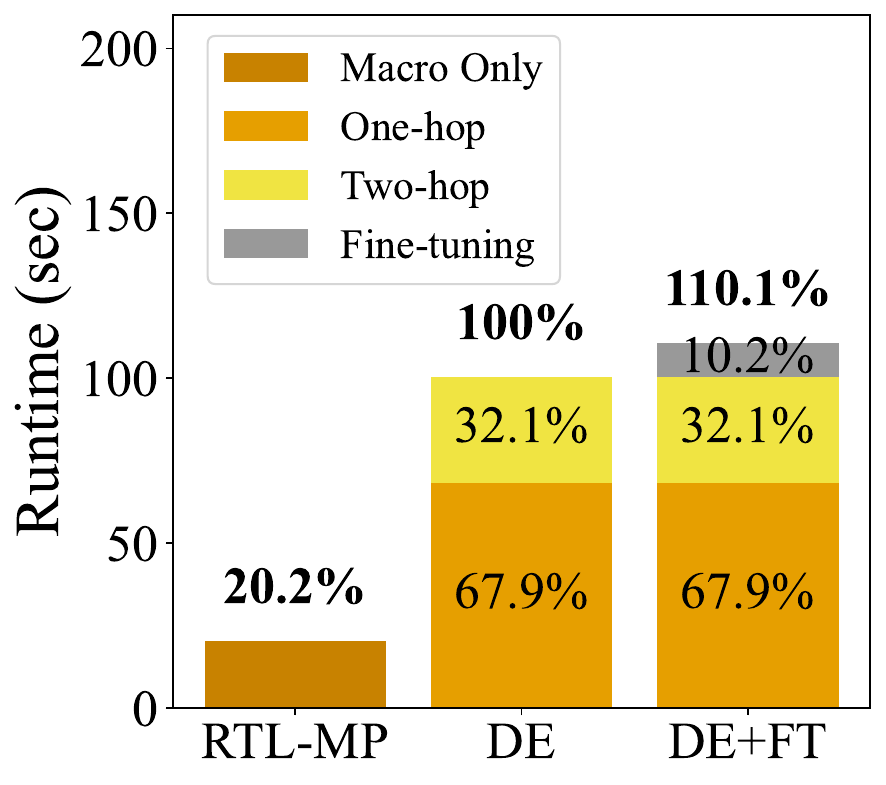}
    \label{fig:runtime2}
    }
    \subfigure[{\tt ariane133} breakdown]{
    \includegraphics[width=0.47\linewidth]{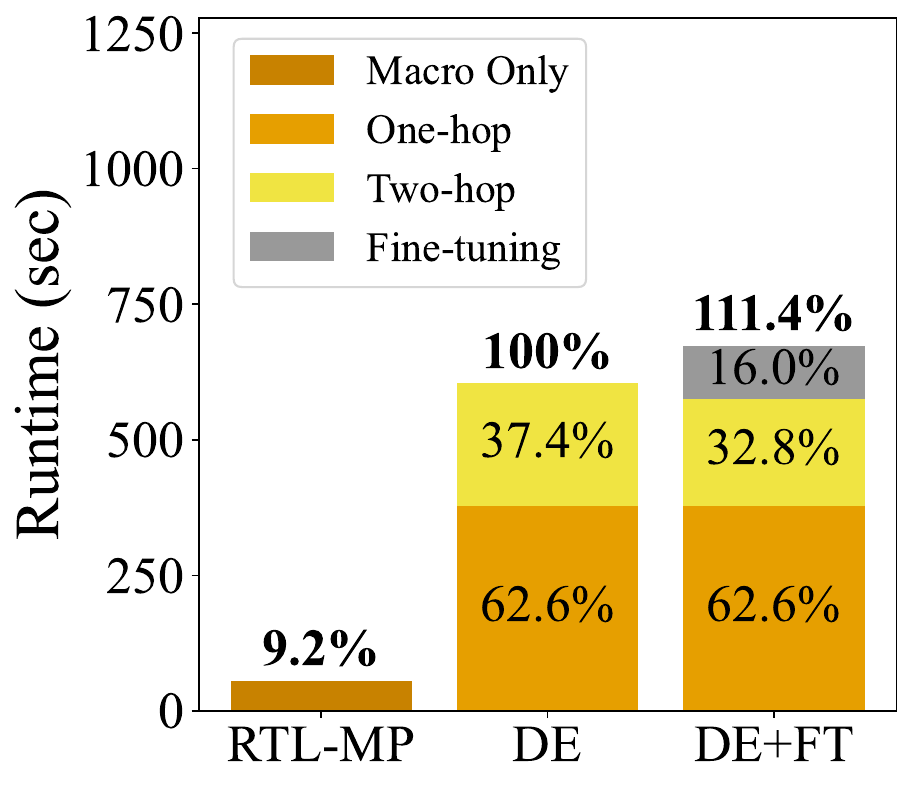}
    \label{fig:runtime3}
    }
  \caption{Runtime breakdown comparison between RTL-MP \cite{Kahng2022}, DAS-MP (DE) \cite{10546560} and DAS-MP (DE+FT) for designs {\tt swerv\_wrapper} (left) and {\tt ariane133} (right). The time spent by each component is normalized to the total runtime of DAS-MP (DE) \cite{10546560}.}
    \label{fig:runtime}
\end{figure}

The runtime of the proposed methodology is compared against RTL-MP \cite{Kahng2022}, with a detailed breakdown summarized in Table \ref{tab:runtime result}. It can be observed that DAS-MP (DE), which includes the {\tt macro cluster-cell cluster} and {\tt macro cluster-cell cluster-cell cluster} connections, results in an average runtime increase of $2.13\times$ and $2.91\times$, respectively. The full methodology, DAS-MP (DE+FT), incorporating optimized {\tt macro cluster-cell cluster-cell cluster} extraction and macro flipping, leads to an average runtime increase of $2.83\times$ and $3.21\times$, respectively. As shown in Fig. \ref{fig:runtime1}, despite the increased runtime in the extraction and fine-tuning phase, the incurred overhead remains less than 1.5\% of the total runtime of the entire macro placement process. Notably, the extraction  and fine-tuning runtime scales sub-linearly with the number of connections, providing significant design quality gains with only a marginal runtime penalty.

For the {\tt macro cluster-cell cluster-cell cluster} extraction within DAS-MP (DE+FT), runtime optimization reduces the overhead to $2.83\times$ due to the introduction of a feedback mechanism. This mechanism not only enhances PPA but also improves runtime efficiency. The acceleration is attributed to the mechanism’s ability to differentiate macros based on area considerations, reducing the overall weight of 2-hop connections and thereby optimizing runtime. Additionally, as illustrated in Fig. \ref{fig:runtime2} and Fig. \ref{fig:runtime3}, we analyzed the runtime of the macro flipping process and found that it accounts for only 10.2\% and 16.0\% of the extraction step. This demonstrates that the flipping action achieves significant PPA optimization with minimal computational cost.

\begin{table*}
\caption{Runtime Analysis}
\centering
\resizebox{1.0\textwidth}{!}{
\begin{threeparttable}
\begin{tabular}{l|c|cc|cc} 
\toprule[1pt]
\multirow{3}{*}{Design Name}    &RTL-MP \cite{Kahng2022}          & \multicolumn{2}{c}{DAS-MP (DE) \cite{10546560} Runtime (s)}        & \multicolumn{2}{c}{DAS-MP (DE+FT) Runtime (s)}       \\
\cmidrule[1pt]{2-6}
                           &Runtime (s) \tnote{1}& Macro Cluster-Cell Cluster \tnote{2} & Macro Cluster-Cell Cluster-Cell Cluster\tnote{3} & Optimized M-C-C Extraction  & Flipping with Dataflow  \\ 
\midrule[1.2pt]
\texttt{TinyRocket}                 &  1.48                   &   6.59         & 9.93 (+3.33)      & 9.91 (\textbf{+3.32})      &  10.22 (+0.31) \\
\texttt{bp\_be}                     &  6.53                   & 17.04          & 23.78 (+6.75)      & 23.66 (\textbf{+6.62})    & 25.64 (+1.98)\\
\texttt{bp\_fe}                     &  1.85                   & 4.49           & 6.11 (\textbf{+1.63})       & 6.13 (+1.64)     &8.28 (+2.15)\\
\texttt{black parrot }              &  326.10                   & 421.98         & 490.42 (+68.44)    & 481.65 (\textbf{+59.67})          &493.63 (+11.94)  \\
\texttt{bp\_multi}                  & 56.29                    &  101.50       &131.52 (+30.02)      & 131.37 (\textbf{+29.88})          &139.57 (+8.20)  \\
\texttt{swerv\_wrapper}             & 20.24                    &    68.14       &100.40 (+32.27)     & 100.38 (\textbf{+32.24})          &110.59 (+10.22) \\  
\texttt{ariane133}                  & 55.48                    &   377.69       & 603.81 (+226.12)    & 576.06 (\textbf{+198.37})          & 673.02 (+96.96)  \\
\midrule[1pt]
\textbf{Average Runtime (s)} & 66.99 (1.00$\times$) & 142.49 (2.13$\times$) & 195.28 (2.91$\times$) & 189.88 (2.83$\times$)  & 208.71 (3.12$\times$) \\
\bottomrule[1.2pt] 
\end{tabular} 
\label{tab:runtime result}
\begin{tablenotes}
        \footnotesize
        \item[1] Extracting macro cluster-macro cluster connections only within RTL-MP \cite{Kahng2022}. 
        \item[2] Includes extracting connections among macro clusters.
        \item[3] Includes extracting previous two types of connections in 1 and 2.
      \end{tablenotes}
\end{threeparttable}}
\vspace{-8pt}
\end{table*}

\begin{table*}[t]
\vspace{-8pt}
\centering
\caption{Ablation Study Results for Fine-tuning Processes}
\vspace{-3pt}
\label{tab:ablation study}
\resizebox{0.83\textwidth}{!}{
\begin{tabular}{@{}l|l|l|l|l|l|l|l@{}}
\toprule[1.5pt]
\textbf{Design}                                                       & \textbf{Type}        & \textbf{HPWL} & \makecell[l]{\textbf{Congestion}\\ \textbf{Overflow}} & \textbf{WNS ($ns$)} & \textbf{TNS ($ns$)} & \textbf{Power ($mW$)} & \textbf{Area ($\mu m^2$)} \\ \midrule
\multirow{5}{*}{\textbf{\texttt{TinyRocket}}}        
& RTL-MP\cite{Kahng2022}           & 803.80                             & 74                                                 & -0.26                                & -83.64                               & 73.4                                    & 59538                          \\
& DAS-MP (DE) \cite{10546560}   & 752.80         &    19            &    -0.23        &      -68.88       &  73.2        &  59539                                \\ 
& DAS-MP (DE+FT) (Area Fine-tuning)  &        \textbf{744.63}   &    \textbf{16 }           & \textbf{-0.22}          & \textbf{-68.53}         &  \textbf{73.2}           &   \textbf{59535 }                              \\ 
& DAS-MP (DE+FT) (Orientation Fine-tuning)  &   746.62       &   \textbf{16}               & \textbf{-0.22}         & \textbf{-68.53}         &   \textbf{73.2}            &  \textbf{59535}                 \\ \midrule
\multirow{5}{*}{\textbf{\texttt{bp\_be}}}                            
& RTL-MP                                              & 4445.60                            & 40537                                              & -0.83                                 & -1005.88                              & 414                                     & 269204   \\
& DAS-MP (DE) \cite{10546560}  & 4113.33         &    454            & -0.62           &            -467.99           &   411  &  268987                           \\ 
& DAS-MP (DE+FT) (Area Fine-tuning)  &  4073.92        &   447             &  \textbf{-0.61}          & -377.34          &   \textbf{389}          &    268990                              \\ 
& DAS-MP (DE+FT) (Orientation Fine-tuning)  & \textbf{4006.38 }       &          \textbf{432 }      &   \textbf{-0.61}  &     \textbf{-318.23 }     &       \textbf{389}        &   \textbf{ 268390 }                              \\\midrule
\multirow{5}{*}{\textbf{\texttt{bp\_fe}}}               
& RTL-MP                                              & 2789.94                           & 4445                                               & -0.70                                & -37.49                               & 264                                     & 221949 \\                    
& DAS-MP (DE) \cite{10546560}  &  2519.63        &    3156            &  -0.53          & -22.33         &   \textbf{259}         &  221774                                \\ 
& DAS-MP (DE+FT) (Area Fine-tuning)  &  2470.9        &  1742              &  -0.52          & -18.95         &    270         &    \textbf{222250}                              \\ 
& DAS-MP (DE+FT) (Orientation Fine-tuning)  & \textbf{2468.93}        &         \textbf{1543}         &     \textbf{-0.49}     &     \textbf{-18.69 }     &      270         &   \textbf{222250 }                             \\\midrule
\multirow{5}{*}{\textbf{\texttt{black parrot}}}                     
 & RTL-MP     & 12930.57      & 981       & -0.14     & -0.15                                & 169                                     & 877590 \\ 
& DAS-MP (DE) \cite{10546560} &  11820.00        &   130             & \textbf{-0.10}           &  \textbf{-0.11}        &   \textbf{154}          & 877477                                 \\ 
& DAS-MP (DE+FT) (Area Fine-tuning)  &  11673.80        &  142              &  \textbf{-0.10}          &  \textbf{-0.11}        & 162            &   \textbf{844700 }                              \\ 
& DAS-MP (DE+FT) (Orientation Fine-tuning)  &  \textbf{11632.06}         &       \textbf{112}        &      -0.11     &     \textbf{-0.11 }       &  169     &    877830                             \\ \midrule
\multirow{5}{*}{\textbf{\texttt{bp\_multi}}}                         
& RTL-MP    & 7252.76   & 3965      & -0.25    & -121.21                              & 218                                     & \textbf{614173}  \\
& DAS-MP (DE) \cite{10546560}  & 7146.48         &    1470            &  \textbf{-0.21}          &  -75.59        & \textbf{207}            &   615342                               \\ 
& DAS-MP (DE+FT) (Area Fine-tuning)  &  7048.47        &  963              & \textbf{-0.21}           & \textbf{-0.21}          &  221           &   614818                               \\ 
& DAS-MP (DE+FT) (Orientation Fine-tuning)  & \textbf{7024.28}        &  \textbf{846}                &  \textbf{-0.21 }     & \textbf{-0.21 }         &    218           & 614231       \\ \midrule
\multirow{5}{*}{\textbf{\texttt{swerv\_wrapper}}}                    

& RTL-MP    & 4745.86     & 2533            & -1.83                                & -3063.16                              & \textbf{26600}                                   & \textbf{511649}  \\
& DAS-MP (DE) \cite{10546560}  & 4648.02         &   \textbf{887}             &  -0.85                   & -1793.16            &  \textbf{26600}      &   511691                        \\ 
& DAS-MP (DE+FT) (Area Fine-tuning)  & 4570.32         &  1053              & -0.85           & -1762.82         &    \textbf{26600}         &  511670                                \\ 
& DAS-MP (DE+FT) (Orientation Fine-tuning)  &  \textbf{4501.14}   & 997    & \textbf{-0.82}     &  \textbf{-1410.25}         & \textbf{26600}     &   511670                                 \\ \midrule
\multirow{5}{*}{\textbf{\texttt{ariana133}}}                        
& RTL-MP     & 8857.12       & 23876             & -0.109    & -2.26                                 & 187                                     & \textbf{739303} \\ 
& DAS-MP (DE) \cite{10546560}                                     & 8624.48                           & 3919                                               & -0.014                                & -0.06                                & \textbf{186}                                     & 739408                                  \\
& DAS-MP (DE+FT) (Area Fine-tuning)  & 8475.60         &  3778              &  -0.014          & -0.06         &  222           &   739309                               \\ 
& DAS-MP (DE+FT) (Orientation Fine-tuning)  &     \textbf{8404.55}      &        \textbf{3678}     &    \textbf{-0.012}      &   \textbf{-0.05}    &      222      &   739310                             \\ \midrule 
\multirow{4}{*}{\textbf{Avg. Improv. (\%)}}  
&RTL-MP (\textbf{Baseline})       & 0.00\%        &   0.00\%          &  0.00\%     &  0.00\%        & 0.00\%       & 0.00\%      \\
& DAS-MP (DE) \cite{10546560}    &  5.46\%                 &      71.49\%        &      36.03\%     &     46.18\%         &    \textbf{2.48\%}       & -0.01\%   \\
& DAS-MP (DE+FT) (Area Fine-tuning)                           & 6.81\%                            & 77.42\%                                            & 36.61\%                               & 56.38\%                               & -1.70\%                                 & \textbf{0.51\%}                                  \\
& DAS-MP (DE+FT) (Orientation Fine-tuning)                      & \textbf{7.28\% }                           & \textbf{79.30\%}                                            & \textbf{38.14\%}                               & \textbf{59.03\%}                               & -2.10\%                                 & 0.01\%   \\                        
\bottomrule[1.5pt]
\end{tabular}
}
\vspace{-8pt}
\end{table*}
\subsection{Ablation Study for Fine-tuning Processes}
To further analyze the impact of different fine-tuning strategies in our macro placement optimization, we conduct an ablation study, summarized in Table \ref{tab:ablation study}, by evaluating two key components: area-based fine-tuning and orientation-based fine-tuning. The table presents quantitative results comparing the baseline RTL-MP, DAS-MP (DE), and the proposed DAS-MP (DE+FT) approaches.

For area-based fine-tuning, our strategy effectively improves congestion and timing metrics. On average, congestion overflow is reduced by 1.43\%, with a maximum improvement of 17.23\% compared to DAS-MP (DE) \cite{10546560}. Timing performance also benefits, achieving an average improvement of 0.58\% in WNS and 10.33\% in TNS, demonstrating the effectiveness of incorporating macro area awareness in placement refinement.

Orientation-based fine-tuning enhances macro placement by leveraging orientation adjustments. As shown in Table \ref{tab:ablation study}, compared to DAS-MP (DE) \cite{10546560}, congestion overflow improves by 2.06\% on average, with a peak improvement of 24.29\%, highlighting the effectiveness of macro orientation refinement. Timing metrics also show notable gains, with WNS and TNS improving by 2.38\% and 17.40\%, respectively. Compared to area-based fine-tuning, orientation-based fine-tuning delivers more consistent improvements, though a slight degradation of 3.06\% is observed in some cases.

These results indicate that both fine-tuning strategies independently contribute to optimization. Their complementary roles underscore the necessity of incorporating both techniques to enhance different aspects of macro placement.

\section{Conclusions}
\label{conclusions}
In this paper, we identified the importance of dataflow analysis in macro placement and proposed DAS-MP, a methodology centered around dataflow extraction and adoption. The methodology first extracts detailed dataflow connections among macro clusters and between macro and cell clusters. These extracted connections are then incorporated into the loss function to guide subsequent macro placement steps. Building upon this foundation, two fine-tuning steps are applied. The first adjusts weights by considering the impact of macro area, aiming to optimize congestion based on engineering practices. Additionally, we recognize the importance of orientation and introduce dataflow-aware approaches to fine-tune macro orientations. Through a diverse set of benchmarks, we demonstrate that the proposed methodology outperforms the recently released academic dataflow-aware macro placer in terms of HPWL and congestion reduction. These improvements are further reflected in significant enhancements in timing results. Moreover, the methodology operates efficiently, with an acceptable overhead that trades computational cost for substantial quality improvements. This work highlights the necessity of detailed and comprehensive dataflow analysis in developing automatic macro placers. It provides an opportunity for co-optimizing macro and standard cell placement, fostering further advancements in placement strategies.
\bibliographystyle{IEEEtran}
\bibliography{ref.bib}

\newpage

 





\end{document}